\documentclass[11pt]{article}

\usepackage{fullpage}
\usepackage{amsmath}
\usepackage{amssymb}
\usepackage{setspace}
\usepackage{bbm}
\usepackage{dsfont}
\usepackage{graphics}

\usepackage[citebordercolor={.8 .8 1},urlbordercolor={.8 .8 1},pdfstartview=FitH]{hyperref}

\usepackage{color}

\usepackage[normalem]{ulem}

\newcommand{\be}{\begin{equation}}
\newcommand{\ee}{\end{equation}}
\newcommand{\bes}{\begin{equation*}}
\newcommand{\ees}{\end{equation*}}
\newcommand{\bea}{\begin{eqnarray}}
\newcommand{\eea}{\end{eqnarray}}
\newcommand{\beas}{\begin{eqnarray*}}
\newcommand{\eeas}{\end{eqnarray*}}

\newcommand{\p}{\partial}

\newcommand{\bmat}{\begin{bmatrix}}
\newcommand{\emat}{\end{bmatrix}}

\def\Tr{{\rm Tr}}
\def\le{\left}
\def\ri{\right}

\def\Ab{\overline{A}}
\def\ab{\overline{a}}

\def\wb{\overline{w}}

\def\Tb{\overline{T}}
\def\Lb{\overline{L}}
\def\zb{\overline{z}}
\def\Wb{\overline{W}}
\def\hb{\overline{h}}
\def\qb{\overline{q}}
\def\rt{\rightarrow}
\def\eps{\epsilon}
\def\Pc{{\cal P}}

\def\Jb{\overline{J}}
\newcommand{\twist}{\sigma}
\newcommand{\atwist}{\tilde{\sigma}}
\def\Tr{{\rm Tr}}
\def\le{\left}
\def\ri{\right}

\def\CW{{\cal W}}

\begin{document}
\numberwithin{equation}{section}
{
\begin{titlepage}
\begin{center}

\hfill \\
\hfill \\
\vskip 0.6in

{\Large \bf Higher Spin Entanglement and ${\cal W}_N$ Conformal Blocks}\\

\vskip 0.4in

{\large Jan de Boer${}^a$, Alejandra Castro${}^a$, Eliot Hijano${}^b$, Juan I. Jottar${}^c$ and Per Kraus${}^b$}\\

\vskip 0.3in

${}^{a}${\it Institute for Theoretical Physics, University of Amsterdam,
Science Park 904, Postbus 94485, 1090 GL Amsterdam, The Netherlands} \vskip .5mm
${}^{b}${\it Department of Physics and Astronomy, University of California, Los Angeles, CA 90095, USA} \vskip .5mm
${}^{c}${\it Institut f\"ur Theoretische Physik, ETH Z\"urich, CH-8093 Z\"urich, Switzerland } \vskip .5mm

\end{center}

\vskip 0.45in

\begin{center} {\bf ABSTRACT } \end{center}

Two-dimensional conformal field theories with extended {$\cal{W}$}-symmetry algebras have dual descriptions in terms of weakly coupled higher spin gravity in AdS$_3\,$ at large central charge. Observables that can be computed and compared in the two descriptions include R\'enyi and entanglement entropies, and correlation functions of local operators. We develop techniques for computing these, in a manner that sheds light on when and why one can expect agreement between such quantities on each side of the duality. We set up the computation of excited state R\'enyi entropies in the bulk in terms of Chern-Simons connections, and show how this directly parallels the CFT computation of correlation functions. More generally, we consider the vacuum conformal block for general operators with $\Delta \sim c\,$. When two of the operators obey ${\Delta \over c} \ll 1\,$, we show by explicit computation that the vacuum conformal block is computed by a bulk Wilson line probing an asymptotically AdS$_3$ background with higher spin fields excited, the latter emerging as the effective bulk description of the excited state produced by the heavy operators. Among other things, this puts a previous proposal for computing higher spin entanglement entropy via Wilson lines on firmer footing, and clarifies its relation to CFT. We also study the corresponding computation in Toda theory and find that this provides yet another independent way to arrive at the same result.

%\vfill
%
%\noindent Draft version of \today

\end{titlepage}
}
%%%%%%%%%%%%%%%%%%%%%%%%%%%%%%%%%%%%%%%%%%%%%%%%%%%%%%%%%%%%%%%%%%

%%%%%%%%%%%%%%%%%%%%%
\newpage

\tableofcontents

%%%%%%%%%%%%%%%%%%%%%%%%%%%%%%%%%%%%%%%%%%%%%%%%%%%%%%%%%%%%%%%%%%%%%%%%%%%%%%%%
\section{Introduction}
%%%%%%%%%%%%%%%%%%%%%%%%%%%%%%%%%%%%%%%%%%%%%%%%%%%%%%%%%%%%%%%%%%%%%%%%%%%%%%%%

Entanglement entropy has emerged as an interesting new observable in quantum field theory, yielding information that goes beyond that provided by correlation functions of local operators. For reviews see e.g. \cite{Nishioka:2009un,Calabrese:2009qy}. It has applications to a diverse set of systems, ranging from condensed matter to string theory and holographic duality. In the latter context, a particularly appealing feature is that entanglement entropy admits a beautifully simple realization via the Ryu-Takayanagi formula \cite{Ryu:2006bv}, indicating a potentially far reaching link between quantum entanglement and spacetime geometry. More generally, one can study the R\'enyi entropy, which captures the full content of the reduced density matrix for a subsystem.

To the extent that the Ryu-Takayanagi formula is an important clue to understanding the mechanism underlying holography, it is of great interest to generalize to theories that go beyond ordinary Einstein gravity. One such direction that has been explored is the case in which higher derivative terms are included in the action \cite{Hung:2011xb,Dong:2013qoa,Camps:2013zua}. Another is to the case of higher spin gravity, which is the case that we focus on here, in particular its 3$d$ version and corresponding 2$d$ CFT dual. The study of entanglement entropy in higher spin theories was initiated in \cite{Ammon:2013hba,deBoer:2013vca} and further work appears in \cite{Chen:2013dxa,Perlmutter:2013paa,Datta:2014ska,Datta:2014uxa,
Castro:2014tta,Datta:2014ypa,Hijano:2014sqa,Castro:2014mza,Bagchi:2014iea}. Our two main goals are the following: first to attempt to derive from first principles the prescriptions advanced in \cite{Ammon:2013hba,deBoer:2013vca},\footnote{These two proposals were recently shown to be equivalent in \cite{Castro:2014mza}.} and second to extend these considerations to the case of R\'enyi entropy in higher spin theories.

In quantum field theory a useful way to compute entanglement entropy is via the replica trick. If $\rho_{A}$ is the reduced density matrix for a subregion $A$, the idea is to compute $\Tr \left[\rho_{A}^n\right] $ by introducing $n$ copies of the original field theory with certain twisted boundary conditions. Equivalently, one computes the path integral for the original theory defined on an $n$-sheeted branched cover of the original spacetime. When the field theory in question is a $2d$ CFT, in the semiclassical (large-$c$) limit one can envision computing the partition function on the branched cover using holography, namely as the saddle point approximation to the partition function of a three-dimensional bulk theory that contains gravity. In the present paper we will focus on CFTs which are dual to $sl(N,\mathds{R})\oplus sl(N,\mathds{R})$ Chern-Simons theories in the bulk.\footnote{More precisely, the Chern-Simons theory captures the chiral algebra of the CFT; additional degrees of freedom are needed to describe the full operator spectrum.} These include pure AdS$_{3}$ gravity ($N=2$) \cite{Achucarro:1989gm,Witten:1988hc}, gravity coupled to Abelian gauge fields, and also higher spin theories which are dual to CFTs with extended symmetry algebras of $\mathcal{W}$-type \cite{Campoleoni:2010zq,Campoleoni:2011hg,Gaberdiel:2011wb}. One of our goals is to describe how to exploit the Chern-Simons description to simplify the calculation of interesting non-local observables in this class of CFTs, and build towards a constructive proof of the holographic higher spin entanglement entropy proposal of \cite{Ammon:2013hba,deBoer:2013vca}.

For the 2$d$ CFTs dual to pure gravity, a holographic computation of vacuum state R\'enyi entropies in the large-$c$ limit was accomplished in \cite{Headrick:2010zt,Faulkner:2013yia}, and independently in \cite{Hartman:2013mia} from a field theory perspective. The holographic calculation in \cite{Headrick:2010zt,Faulkner:2013yia} is based on Schottky uniformization, wherein the replica Riemann surface is described as a quotient of the complex plane by a discrete subgroup $\Sigma$ of PSL$(2,\mathds{C})$ (see e.g. \cite{Zograf:1988}). The bulk manifolds corresponding to these boundary topologies are handlebodies which can be described as a quotients of AdS$_{3}$ by $\Sigma$ \cite{Krasnov:2000zq}. Roughly speaking, these gravitational saddles correspond to particular ways of ``filling in" the boundary Riemann surface with Euclidean AdS$_{3}$ space. These observations were used in \cite{Headrick:2010zt,Faulkner:2013yia} in order to evaluate the regularized gravitational action on the handlebody solutions, thus computing the large-$c$ R\'enyi entropies for a subsystem consisting of disjoint intervals in the dual $2d$ CFT.

In Chern-Simons language, the task at hand consists of constructing the flat connections that are compatible with the bulk replica Riemann surface. This perspective provides a convenient way of organizing the calculation of R\'enyi and entanglement entropies which circumvents some of the complications associated with the use of metric variables. But, more importantly, the Chern-Simons formulation generalizes straightforwardly to a class of theories beyond pure gravity, such as the higher spin theories mentioned above, and as we shall see accommodates the case of excited states without too much difficulty. Furthermore, while still technically challenging in practice, the topological formulation offers an a priori systematic way of deforming the CFT by incorporating sources for the stress tensor and other conserved currents.

We now spell out a few more details about our approach.
Our treatment of the Chern-Simons theory will be entirely classical, and as such is only valid in the limit of large central charge. We wish to obtain results for the R\'enyi entropy in excited states which have a nice classical limit as nontrivial deformations of AdS$_3\,$; this requires that the energy and higher spin charges of the background scale like the central charge. By the state-operator map, the corresponding operators that create such excited states have quantum numbers that inherit this scaling. Furthermore, it is standard to think of the branched cover as being created by the insertion of twist operators that sew together distinct copies of the replica theory, and these twist operators have conformal dimensions that also scale like the central charge.

An attractive aspect of the Chern-Simons formulation of this problem is that it leads to equations that directly match up to those obtained in a purely CFT analysis of correlation functions in this semiclassical limit. This is also the case in the metric formulation, but here the connection is more immediate and transparent. More precisely, decomposing such correlation functions into conformal blocks, the conformal blocks are obtained by solving a certain monodromy problem. This monodromy problem is the same one that is encountered upon demanding that the Chern-Simons connection has the correct holonomies dictated by the branched covering. It then becomes clear that what the Chern-Simons action is computing is a conformal block: the contribution to the correlation function due to intermediate states which are (Virasoro or more generally ${\cal W}_N$) descendants of a primary state. If one can argue that a specific block dominates in the semiclassical limit, one thereby establishes that the Chern-Simons computation indeed yields the correct R\'enyi entropy. This line of reasoning follows that of \cite{Faulkner:2013yia,Hartman:2013mia} in the case of ordinary gravity, but now formulated more efficiently in Chern-Simons language.

%\textcolor{red}{\sout{the vacuum}}

In general it is difficult in practice to compute the R\'enyi entropy in an excited state by this method, because it requires the solution of a differential equation that is typically intractable. Things simplify greatly if one focusses instead on entanglement entropy. The key point here is that this involves taking the replica index $n\rightarrow 1$, and in this limit the dimension of the twist operator goes to zero. One just needs to solve the monodromy problem to first order in $n-1$, which is quite straightforward. In the case of ordinary gravity, one thereby derives the Ryu-Takayanagi formula for the entanglement entropy in an excited state, where the bulk description of the excited state is as a conical defect or BTZ solution \cite{Fitzpatrick:2014vua,Asplund:2014coa}.

When higher spin fields are turned on in the bulk, the Ryu-Takayanagi prescription no longer applies; it has been proposed \cite{Ammon:2013hba,deBoer:2013vca} that one should instead evaluate a certain Wilson line observable, which is indeed a rather natural object in the Chern-Simons formulation. We want to establish the validity of this proposal. Focussing on the case of $sl(3,\mathds{R})\oplus sl(3,\mathds{R})$ Chern-Simons theory, what we will show by explicit computation is that the Wilson line evaluated in a general asymptotically AdS$_3$ background computes, in a rather efficient fashion, the same answer as that produced by the monodromy analysis. We therefore find that the Wilson line yields the ${\cal{W}}_3$ vacuum block contribution to the correlation function of two twist operators and two excited state operators. As noted above, whether this gives the entanglement entropy hinges on whether the vacuum block is the dominant contribution in the semiclassical limit, but this requires specifying more information about the precise CFT under consideration. The same issue is present in the case of ordinary gravity. That is to say, the validity of the Wilson line proposal for computing entanglement entropy in an excited state of the higher spin theory is now on the same footing as the validity of the Ryu-Takayanagi formula for an excited state in ordinary gravity.

Actually, when phrased in terms of conformal blocks the problem of computing R\'enyi entropy in the Chern-Simons formulation is a special case of a more general problem of interest. Namely, we can replace the twist operators by general operators that carry both a scaling dimension and a higher spin charge, and think of computing the corresponding vacuum block. This more general setup again can be formulated as a monodromy problem, both in Chern-Simons theory and in the CFT. And we again can demonstrate agreement between the perturbative solution of the monodromy problem and the result produced by a general Wilson line observable. This provides a satisfying answer to the question of what the general Wilson line evaluated in the general higher spin background is computing: it is computing the ${\cal W}_N$ vacuum block contribution to the four-point function, where two of the operators correspond to the background solution and the other two to the Wilson line.

An independent way to approach the computation of R\'enyi entropy in the Chern-Simons formulation is to first integrate out the bulk
fields and to phrase everything in terms of an effective theory on the boundary. This effective theory is Liouville theory for
pure gravity, and Toda theory for the higher spin case, and our computations are related to four-point functions in these theories. 
As we will demonstrate, the relevant four-point functions are essentially fixed by the symmetries of the problem, and this provides
yet another interesting perspective on the problem.

The remainder of this paper is organized as follows. In section \ref{sec:renyi} we briefly review the computation of R\'enyi entropy via correlation functions of twist operators. We then discuss the conformal block decomposition of such correlators in section \ref{sec:mondromy}, and show how to compute the vacuum block in the semiclassical limit by solving a monodromy problem. The corresponding bulk problem in Chern-Simons language is discussed in section \ref{bulkmono}. In section \ref{sec:pert} we discuss the perturbative solution of the monodromy problem.   We then compute the Wilson line in section \ref{wilson} and demonstrate agreement with the result obtained from the monodromy analysis.  Another perspective based on Toda field theory is discussed in section \ref{toda}.  In this approach one obtains correlation functions by a saddle point approximation to the Toda path integral.   We show how to connect the Toda field at the saddle point to the data in the Chern-Simons construction.     We discuss some aspects of our results in section \ref{discussion}, including the emergence of black hole solutions as effective descriptions of CFT microstates. The appendices collect useful formulas and conventions, and some complementary material.

%%%%%%%%%%%%%%%%%%%%%%%%%%%%%%%%%%%%%%%%%%%%%%%%%%%%%%%%%%%%%%%%%%%%%%%%%%%%%%%%
\section{R\'enyi entropies}\label{sec:renyi}
%%%%%%%%%%%%%%%%%%%%%%%%%%%%%%%%%%%%%%%%%%%%%%%%%%%%%%%%%%%%%%%%%%%%%%%%%%%%%%%%

The standard way of computing entanglement and R\'enyi entropies in quantum field theory is based on the replica trick: given a constant-time region $A\,$, the R\'enyi entropies are defined as
\begin{equation}
S_{A}^{(n)} = -\frac{1}{n-1}\ln \text{Tr}\bigl[\rho_{A}^{n}\bigr]~\qquad n \geq 2\,,
\end{equation}
\noindent where $\rho_{A} =\text{Tr}_{B}\left[\rho\right]$ is the reduced density matrix associated with region $A\,$, and $B=A^c\,$. In particular, assuming a unique analytic continuation in $n$ exists, the entanglement entropy $S_{A}$ for subsystem $A$ is then obtained as
\begin{equation}
S_{A} = - \text{Tr}\bigl[\rho_{A}\log \rho_{A}\bigr]= \lim_{n\to 1}S_{A}^{(n)}\,.
\end{equation}

In a two-dimensional QFT, in the case where $A$ is the union of $N_{I}$ disjoint intervals one finds (see e.g. \cite{Calabrese:2004eu,Cardy:2007mb,Calabrese:2009qy})
\begin{equation}\label{Renyi from partition function}
S_{A}^{(n)} = -\frac{1}{n-1}\Bigl[\ln Z(\mathcal{R}_{n,N_{I}}) - n \ln Z_{1}\Bigr],
\end{equation}
\noindent where $Z$ denotes the partition function and $\mathcal{R}_{n,N_{I}}$ is a Riemann surface obtained by cutting the spacetime (a cylinder or plane, say) along $A$ and cyclically gluing $n$-copies along the cut, with the last copy joined to the first. $Z_{1}$ denotes the partition function of the theory on the original manifold, with no branch points. If the theory was originally defined on the plane (or the cylinder), the corresponding replica manifold is a Riemann surface of genus
\begin{equation}
g(\mathcal{R}_{n,N_{I}}) = (n-1)(N_{I}-1)~,
\end{equation}
\noindent with $2N_{I}$ branch points $z_{i}\,$, which can be defined by the curve \cite{Faulkner:2013yia}
\begin{equation}
y^{n} = \prod_{i=1}^{N_{I}}\frac{z-z_{2i-1}}{z-z_{2i}}\,.
\end{equation}

An alternative description is obtained by replacing the original field theory defined on the $n$-sheeted surface by $n$ copies of the field theory defined on the original surface. Going around a branch point permutes the copies in a manner that reproduces the $n$-sheeted construction. In this picture, the branch points correspond to the location of twist operators. In a conformal field theory, these are primaries of dimension $(\Delta,\Delta)$ with $\Delta={c\over 24}(n-{1\over n})\,$. In this language we have
\be
Z(\mathcal{R}_{n,N_{I}}) = \bigl\langle \twist(z_1,\overline{z}_1)\atwist(z_2,\overline{z}_2)\ldots \twist(z_{2N_I-1},\overline{z}_{2N_I-1})\atwist(z_{2N_I},\overline{z}_{2N_I}) \bigr\rangle
\ee
where $\twist$ and $\atwist$ denote twist and anti-twist operators. This construction yields the R\'enyi entropy in the vacuum state, but we can generalize to an excited state as follows \cite{Nozaki:2014hna,Caputa:2014vaa}. Consider an excited state ${\cal O}(z,\bar{z})|0\rangle$ obtained by acting with a primary operator on the vacuum state. The reduced density matrix $\rho_A$ is now obtained by doing a path integral with insertions of $ {\cal O}(z,\bar{z}) $ and $[{\cal O}(z,\bar{z})]^*$ representing the initial and final excited states. Passing to the $n$-sheeted geometry $\mathcal{R}_{n,N_{I}}$ to compute $\Tr \left[\rho_A^n\right]\,$, we now have insertions of these local operators on every sheet. Passing finally to the replica theory defined on the original surface, we end up with an insertion of operators that are the product of $n$ local operators, one from each replica copy. To summarize, in an excited state created by a local operator we need to compute the $2N_I+2$ point correlation function
\be
\label{excited Z}
Z_{\cal O}(\mathcal{R}_{n,N_{I}}) = \left\langle \twist(z_1,\overline{z}_1)\atwist(z_2,\overline{z}_2)\ldots \twist(z_{2N_I-1},\overline{z}_{2N_I-1})\atwist(z_{2N_I},\overline{z}_{2N_I}){\cal O}^{(n)}(z,\bar{z})[{\cal O}^{(n)}(z,\bar{z})]^* \right\rangle
\ee
where
\be
{\cal O}^{(n)}(z,\bar{z}) = \prod_{a=1}^n {\cal O}_a(z,\bar{z})~,
\ee
with $a$ being the replica index.

We will focus primarily on the case of a single interval, $N_I=1\,$, in which case (\ref{excited Z}) is a four-point function of primary operators. In general, computing this correlation function for arbitrary $n$ is as hard as computing a $2n$ point function of the operator ${\cal O}$ in the original CFT; indeed, one obtains such a description by mapping the surface $\mathcal{R}_{n,1}$ to the plane. However, simplifications occur in the semi-classical limit in which the central charge $c$ is taken to infinity. Let us discuss the generic case of a four-point function of primaries,
\be
G(x_1,x_2,x_3,x_4)=\bigl\langle {\cal O}_1 (x_1) {\cal O}_2 (x_2) {\cal O}_3 (x_3) {\cal O}_4 (x_4) \bigr\rangle~.
\ee
In the basic case in which we just have Virasoro symmetry, the semiclassical limit is obtained by considering operators whose conformal dimensions $\Delta$ grow like $c\,$, so that ${\Delta \over c}$ is held fixed as $c\rightarrow \infty\,$. As we recall in more detail in the next section, the four-point function can be decomposed into conformal blocks. The conformal blocks only depend on operator dimensions and the central charge, and essentially capture all of the information imposed by conformal symmetry. In favorable circumstances, the lowest dimension conformal block corresponding to the exchange of the identity operator and its Virasoro descendants will dominate in the semiclassical limit. The problem then reduces to computing the Virasoro identity block in the semiclassical limit, which is a much simpler problem, and one that reduces to computing the monodromies of a certain differential equation, as we discuss in the next section.

It is easy to come up with examples for which the correlation function is not dominated by the Virasoro identity block. A case that is highly relevant for our purposes is the following. If the symmetry algebra of the theory contains higher spin currents in the form a ${\cal W}_N$ algebra, then primaries are labelled by their higher spin charges, along with their conformal dimension. If the higher spin charges $Q$ grow with $c$ such that ${Q\over c}$ is held fixed in the semiclassical limit, then there is no reason to expect that the Virasoro blocks corresponding to the exchange of the higher spin currents and their descendants are suppressed relative to the identity block. But one can still hope for a simplification, namely that what dominates is the identity block of the full ${\cal W}_N$ algebra. The ${\cal W}_N$ identity block contains the Virasoro identity block along with the Virasoro blocks of all operators
constructed from the higher spin currents.

Thus, our working assumption will be that the R\'enyi entropy for excited states containing higher spin charges can be obtained from the vacuum block of the ${\cal W}_N$ algebra in the semiclassical limit. We therefore need to generalize the method that yield the semiclassical Virasoro blocks to the ${\cal W}_N$ context. We do this in the next section, focussing on the case of ${\cal W}_3\,$.

%%%%%%%%%%%%%%%%%%%%%%%%%%%%%%%%%%%%%%%%%%%%%%%%%%%%%%%%%%%%%%%%%%%%%%%%%%%%%%%%
\section{The monodromy problem from CFT }\label{sec:mondromy}
%%%%%%%%%%%%%%%%%%%%%%%%%%%%%%%%%%%%%%%%%%%%%%%%%%%%%%%%%%%%%%%%%%%%%%%%%%%%%%%%

In this section we first provide a very brief review of the monodromy method for computing Virasoro conformal blocks in the semiclassical limit, and then explain how this method works for the ${\cal W}_3$ algebra. Our discussion of the Virasoro blocks follows closely \cite{Harlow:2011ny,Hartman:2013mia,Fitzpatrick:2014vua}.

We first define conformal blocks by inserting the identity in the four point function and expanding it as a sum over a complete set of states $|\chi\rangle$
\be
\bigl\langle {\cal O}_1 (x_1) {\cal O}_2 (x_2) {\cal O}_3 (x_3) {\cal O}_4 (x_4)\bigr \rangle=\sum_{\chi}\bigl\langle {\cal O}_1 (x_1) {\cal O}_2 (x_2)\bigl | \chi\bigr\rangle \bigl\langle \chi \bigl | {\cal O}_3 (x_3) {\cal O}_4 (x_4) \bigr\rangle ~.
\ee
In a CFT, the Hilbert space can be organized into irreducible representations of the Virasoro algebra, with each such representation being labelled by a primary state $|\alpha\rangle\,$. Being somewhat schematic and now letting $|\alpha\rangle $ stand for the primary state and all of its descendants, we denote the contribution of a single Virasoro representation as ${\cal F}_\alpha\,$, so that
\be
\bigl\langle {\cal O}_1 (x_1) {\cal O}_2 (x_2) {\cal O}_3 (x_3) {\cal O}_4 (x_4) \bigr\rangle= \sum_\alpha \bigl\langle {\cal O}_1 (x_1) {\cal O}_2 (x_2) \bigl | \alpha \bigr\rangle \bigl\langle \alpha \bigl | {\cal O}_3 (x_3) {\cal O}_4 (x_4) \bigl\rangle \equiv \sum_\alpha {\cal F}_\alpha(x_i)~.
\ee
${\cal F}_\alpha$ is the conformal partial wave associated to a given Virasoro representation $\alpha\,$.

We now define the semiclassical limit as the limit $\Delta, c\rightarrow \infty$ with the ratio $\Delta/c$ kept finite, where $\Delta$ refers to the conformal dimensions of the external operators as well as the primary $\alpha\,$. In this limit it is expected that the conformal blocks exponentiate \cite{Zamo:1986,Harlow:2011ny}
\be
\bigl\langle {\cal O}_1 (x_1) {\cal O}_2 (x_2) \bigl | \alpha \bigr\rangle \bigl\langle \alpha \bigl | {\cal O}_3 (x_3) {\cal O}_4 (x_4) \bigr\rangle= {\cal F}_{\alpha}(x_i)\approx e^{-{{c}\over{6}} f(x_i)} ~,
\ee
where the function $f(x_i)$ depends on $\Delta$ and $c$ only through the ratio $\Delta/c\,$.
There is no rigorous derivation of this statement, but a considerable amount of evidence in its favor has accumulated \cite{Zamolodchikov:1995aa}. Here we will assume this to be correct. The next step is to insert an operator $\hat \psi(z)$ whose dimension is held fixed in the semiclassical limit. The argument to be made is that the conformal block gets multiplied by a wave function $\psi(z,x_i)$
\be\label{Psidef}
\Psi(z,x_i)=
\bigl\langle {\cal O}_1 (x_1) {\cal O}_2 (x_2)\bigl | \alpha \bigr\rangle \bigl\langle \alpha \bigl | {\hat\psi(z)} {\cal O}_3 (x_3) {\cal O}_4 (x_4) \bigr\rangle= \psi(z,x_i){\cal F}_{\alpha}(x_i) ~.
\ee
This can be seen as the definition of $\psi(z,x_i)\,$. The crucial property is that $\psi$ and its derivatives are of order ${\cal O}(e^{c^0})\,$. This is very powerful, as $\hat \psi(z)$ can be chosen to be a degenerate operator of the theory. The shortening condition of the degenerate operator imposes a differential equation on the wave function $\psi(z,x_i)\,$.

It will prove convenient to keep the general notation $\mathcal{O}_{i}$ with $i=1,2,3,4$ for the time being; in the end, however, we will take ${\cal O}_3$ and ${\cal O}_4$ to be heavy operators with quantum numbers denoted by subindex $2$, and ${\cal O}_1$ and ${\cal O}_2$ to be operators that become light upon analytic continuation $n \to 1$ in the replica number, with quantum numbers denoted by subindex $1$. In a slightly unorthodox nomenclature, from now on we will refer to the latter as ``light operators", but it should be kept in mind that their dimensions scale as $\Delta \sim \mathcal{O}(c)$ with the central charge, which is a crucial requirement for the exponentiation of the conformal block. More concretely, our nomenclature in the reminder of the paper will be
\begin{alignat}{2}\label{eq:lh}
\text{``light":}&\qquad &
\frac{\Delta}{c}
={}&
\mathcal{O}(n-1)
\\
\text{``heavy":}&\qquad &
\frac{\Delta}{c}
={}&
\mathcal{O}(1)
\end{alignat}

\noindent The light operators $\mathcal{O}_{1}$, $\mathcal{O}_{2}$ (typically twist operators) will be located at $1$ and $x$, while the heavy operators $\mathcal{O}_{3}$, $\mathcal{O}_{4}$ (typically creating the excited state) will be inserted at $0$ and $\infty\,$.

We now turn to the study of the consequences of inserting the light operator $\hat\psi$ in a CFT with Virasoro or ${\cal W}_3$ symmetry.

%%%%%%%%%%%%%%%%%%%%%%%%%%%%%%%%%%%%%%%%%%%%%%%%%%%%%
\subsection{Virasoro algebra}\label{sec:cftsl2}
%%%%%%%%%%%%%%%%%%%%%%%%%%%%%%%%%%%%%%%%%%%%%%%%%%%%%

In the case of the Virasoro algebra we can find a primary with the following shortening condition
\be
\left(L_{-2}-{{3}\over{2(2\Delta_{\psi}+1)}}L_{-1}^2\right)\bigl |\hat \psi \bigr\rangle=0 ~,
\ee
provided $\Delta_\psi = {1\over 16} \left[5-c\pm \sqrt{(c-1)(c-25)}\right]\,$, which means that the corresponding representation of the Virasoro algebra contains a null vector at level two. Choosing the $+$ sign in the definition of $\Delta_\psi\,$, we have $\Delta_{\psi}\rightarrow- {1\over 2} -{9\over 2c}$ in the semi-classical limit, so that the shortening condition reads
\be\label{eq:sc1}
\left(L_{-2}+{{c}\over{6}}L_{-1}^2\right)\bigl|\hat \psi \bigr\rangle=0 ~.
\ee
Acting with this condition on $\hat\psi$ inside $\Psi(z,x_i)$ as defined in (\ref{Psidef}) implies the following differential equation in the $z$ variable
\be\label{eq:T1}
\psi''(z)+T(z)\psi(z)=0 ~,
\ee
where $T(z)$ is given by
\be \label{jjj1}
T(z)=
\frac{
\bigl\langle \hat{T}(z) {\cal O}_1 (x_1) {\cal O}_2 (x_2) {\cal O}_3 (x_3) {\cal O}_4 (x_4) \bigr\rangle}{
\bigl\langle {\cal O}_1 (x_1) {\cal O}_2 (x_2) {\cal O}_3 (x_3) {\cal O}_4 (x_4) \bigr\rangle}=
\sum_i \left({{h_i}\over{(z-x_i)^2}}+{{c_i}\over{(z-x_i)}}\right),
\ee
\noindent where $\hat{T}(z)$ denotes the stress tensor as an operator. Here
\be
h_i= {6\over c}\Delta_i~,
\ee
are the rescaled conformal dimensions of the operators ${\cal O}(x_i)\,$, and $c_i$ are auxiliary parameters related to the conformal blocks $f(x_i)$ through a derivative
\be
c_i\equiv {\partial f\over \partial x_i}~.
\ee
Three of these auxiliary parameters can be fixed by demanding smoothness of $T(z)$ as $z\rightarrow\infty\,$, which requires the large $z$ falloff $T(z)\sim {\cal O}(z^{-4})\,$.
Demanding this and sending ($x_1,x_2,x_3,x_4$) to ($1,x,0,\infty$) with a global conformal transformation implies\footnote{Recall that $h_1$ denotes the chiral conformal dimension of the operators at $x$ and $1$; while $h_2$ refers to the operators at $0$ and $\infty\,$.}
\be \label{tparam}
T(z)={{h_2}\over{z^2}}+\left({{1}\over{(z-1)^2}}+{{1}\over{(z-x)^2}}+{{2}\over{(1-z)z}}\right)h_1+c_x {{x(1-x)}\over{z(1-z)(z-x)}}~.
\ee
Equivalently, this follows from (\ref{jjj1}) by writing the most general conformally invariant four-point function
in terms of the standard cross-ratios of the $x_i$ and using the Ward identity for the energy-momentum tensor. Either way,
we then have a differential equation for $\psi(z)$ that involves the parameter $c_x\,$. We need some constraint on the solutions of this differential equation to obtain $c_x\,$. It turns out that the family of solutions must have specific monodromy, and this arises again from the degeneracy of $\hat\psi\,$. To see this we look at the OPE between ${\cal O}_3(x_3){\cal O}_4(x_4)$ inside the $\langle \alpha |\hat\psi(z) {\cal O}_3{\cal O}_4\rangle$ part of the conformal block. This results in a sum of three point functions
\be
\sum_{\beta}c_{34\beta}\bigl\langle \alpha \bigr|{\hat \psi}(z) {\cal O}_\beta\bigr\rangle~.
\ee
Applying the shortening condition \eqref{eq:sc1} on this expression imposes a constraint on $h_{\beta}$ that restricts its value to two possibilities for each choice of $\alpha\,$. This implies that moving $\hat\psi(z)$ around $x_3$ and $x_4$ must have monodromy consistent with these values of $h_{\beta}\,$. As we are studying the $\alpha$-conformal block in the s-channel, this means that $\hat\psi(z)$ must have the same monodromy when moving around $x_1$ and $x_2\,$. For the case of the identity block, it turns out that the monodromy must be the identity.

%%%%%%%%%%%%%%%%%%%%%%%%%%%%%%%%%%%%%%%%%%%%%%%%%%%%%
\subsection{${\cal W}_3$ algebra}\label{sec:block3}
%%%%%%%%%%%%%%%%%%%%%%%%%%%%%%%%%%%%%%%%%%%%%%%%%%%%%

We now turn to the study of the degenerate operators of the ${\cal W}_3$ algebra,
\bea\label{W3 algebra commutators}
[L_m,L_n]&=&(m-n)L_{m+n}+{c\over{12}}m(m^2-1)\delta_{m+n,0}~,\cr
[L_m,W_n]&=&(2m-n)W_{m+n}~,\cr
[W_m,W_n]&=&-{{1}\over{12}}(m-n)(2m^2+2n^2-mn-8)L_{m+n}+{40\over 22+5c}(m-n)\lambda_{m+n}\cr
&&+{{5c}\over{6}}{1\over {5!}}m(m^2-1)(m^2-4)\delta_{m+n,0}~,
\eea
with
\bea
\lambda_{m}=\sum_{n}:L_nL_{m-n}:-{3\over{10}}(m+3)(m+2)L_m~.
\eea
This is a nonlinear algebra on account of the $\lambda_{m+n}$ term. However, we will restrict attention to the semiclassical large-$c$ limit, in which case these terms are suppressed.

It turns out there is a ${\cal W}_3$-primary with null descendants at levels one, two and three. These null states are obtained by evaluating the matrix of inner products among all states at these levels, using the above commutation relations in the large $c$ limit. The operator's quantum numbers in the semiclassical limit are $\Delta_{\psi}=-1$ and $Q_{\psi}=\pm 1/3\,$. The null states are
\bea\label{Wshort}
\left( W_{-1} +{{1}\over{2}}L_{-1}\right)\bigl |\hat\psi \bigr\rangle&=&0 ~, \cr
\left( W_{-2}-L_{-1}^2-{{16}\over{c}}L_{-2} \right)\bigl |\hat\psi \bigr\rangle&=&0 ~, \cr
\left( {{12}\over{c}}L_{-3} -{{24}\over{c}}W_{-3}+{{24}\over{c}}L_{-2}L_{-1}+L_{-1}^3 \right)\bigl |\hat\psi \bigr\rangle&=&0 ~,
\eea
where we have used the first and the second conditions to replace the generators $W_{-1}$ and $W_{-2}$ by Virasoro generators. The conditions listed in (\ref{Wshort}) are only valid in the large $c$ limit, where the nonlinear terms in the ${\cal W}_3$ algebra are suppressed. Inserting this light operator in $\Psi(z,x_i)$ implies the following differential equation in the $z$ variable
\be\label{DE}
\psi'''(z) + 4 T(z) \psi'(z)+2T'(z)\psi(z)-4W(z) \psi(z) =0~.
\ee
The functions $T(z)$ and $W(z)$ come from the insertions of the generators $L_{-2}$ and $W_{-3}$ respectively and they are given explicitly by
\bea
T(z)&=&\sum_i \left({{h_i}\over{(z-x_i)^2}}+{{c_i}\over{(z-x_i)}}\right)~, \cr
W(z)&=&\sum_i \left({{q_i}\over{(z-x_i)^3}}+{{c}\over{6}}{{a_i}\over{(z-x_i)^2}}+{{b_i}\over{(z-x_i)}}\right)~.
\eea
Here as before, $c_i=\partial_i f\,$. $h_i$ and $q_i$ denote the rescaled chiral conformal dimensions and spin-3 charge of the primary at $x_i\,$: $h={6\over c}\Delta$ and $q={6\over c}Q\,$. Smoothness at infinity now also implies $W(z) \sim {\cal O}(z^{-6})$ at large $z\,$. After satisfying these constraints and performing the global conformal transformation to move the operators at ($x_1,x_2,x_3,x_4$) to ($1,x,0,\infty$), these functions read
\bea
T(z)&=&{{h_2}\over{z^2}}+\left({{1}\over{(z-1)^2}}+{{1}\over{(z-x)^2}}+{{2}\over{(1-z)z}}\right)h_1+c_x {{x(1-x)}\over{z(1-z)(z-x)}}~,\cr
W(z)&=&{{q_2}\over{z^3}}+\left({{1}\over{(z-x)^3}}-{{1}\over{(z-1)^3}}\right)q_1\cr
&&+\left(a_1 {{1-x}\over{(z-x)(z-1)^2z}}+a_0 {{x}\over{(z-1)(z-x)z^2}}+a_{x}{{x(1-x)}\over{(1-z)(z-x)^2z}}\right)~.
\eea
We are now using $q_{1,2}$ to denote the spin-3 charges; the operators at $1$ and $x$ carry spin-3 charge $\pm q_1\,$, while those at $0$ and $\infty$ carry $\pm q_2\,$.
The monodromy constraint works in the same way as in the previous subsection. Imposing that the family of solutions of (\ref{DE}) has trivial monodromy around $x_1$ and $x_2$ fixes $a_1$, $a_0$, $a_x$ and $c_x\,$, which we can use to obtain $f(x)$ and ultimately calculate the identity block of the ${\cal W}_3$ algebra.

The structure described above extends in a natural way to the ${\cal W}_N$ case. In this case we will arrive at an $N$th order differential equation, coming from the existence of null states at levels $1$ through $N$, see the discussion below
(\ref{eq:ode3}).

%%%%%%%%%%%%%%%%%%%%%%%%%%%%%%%%%%%%%%%%%%%%%%%%%%%%%%%%%%%%%%%%%%%%%%%%%%%%%%%%
\section{The monodromy problem in the bulk}\label{bulkmono}
%%%%%%%%%%%%%%%%%%%%%%%%%%%%%%%%%%%%%%%%%%%%%%%%%%%%%%%%%%%%%%%%%%%%%%%%%%%%%%%%
A holographic computation of R\'enyi entropies in the semiclassical limit requires evaluating the gravitational action on the appropriate bulk geometry \cite{Headrick:2010zt,Faulkner:2013yia}. The bulk manifolds, while familiar to many, can be rather cumbersome to describe. In this section we will use Chern-Simons language as a convenient way to organize the computation. This route not only circumvents some of the complications associated to metric variables, as we shall see, but it also makes direct contact with the CFT and generalizes straightforwardly to higher spin theories.

We will focus on the three-dimensional $sl(N,\mathds{R})\oplus sl(N,\mathds{R})$ Chern-Simons theory with action
\begin{align}\label{sl(N) CS action}
I_{\rm CS} \equiv{}&
\frac{k_{\rm cs}}{4\pi}\int_{M} \mbox{Tr}\Bigl[CS(A) - CS(\overline{A})\Bigr]~.
\end{align}
The precise field content of the bulk depends on the choice of how the gravitational $sl(2,\mathds{R})$ factor is embedded into $sl(N,\mathds{R})$. In particular, the Chern-Simons level is related to the central charge in the dual theory by
\begin{equation}
k_{\rm cs} = \frac{\ell}{8G_{3}\text{Tr}\left[L_0L_0\right]} = \frac{c}{12\text{Tr}\left[L_0L_0\right]}\,,
\end{equation}
where $\ell$ is the AdS$_{3}$ radius and $L_0$ is the Cartan generator of the $sl(2)$ subalgebra singled out by the choice of embedding.\footnote{We follow the conventions of \cite{Castro:2011iw} for the $sl(N)$ generators.} For concreteness, we will mostly focus on the so-called principal embedding, characterized by the fact that the fundamental representation of $sl(N,\mathds{R})$ becomes an irreducible $sl(2,\mathds{R})$ representation. The resulting bulk theory describes the non-linear interactions of the metric and symmetric tensor fields of spins $s=3,\ldots,N$. % (e.g. $\phi_{\mu\nu\rho} \sim \text{Tr}\left[e_{(\mu}e_{\nu}e_{\rho)}\right]$ and so forth).

On the bulk manifold $M$, let us introduce a radial coordinate $\rho$ and complex coordinates $(z,\bar{z})$ on the $\rho = const.$ slices, which we assume have the topology of the plane or a branched cover thereof. It is convenient to use the gauge freedom of Chern-Simons theory to gauge-away the radial dependence of the connection as
\begin{equation}\label{DS connections on the plane}
A = b^{-1}(\rho)\Bigl(a(z,\zb) + d\Bigr)b(\rho)\,,\qquad \Ab = b(\rho)\Bigl(\bar{a}(z,\zb) + d\Bigr)b^{-1}(\rho)~,
\end{equation}
and concentrate on the ``boundary connections" $a$ and $\bar{a}\,$. Boundary conditions are incorporated by writing the boundary connections in ``Drinfeld-Sokolov" form
\begin{equation}\label{DS connections}
a = \Bigl(L_{1} + T(z)L_{-1} + \sum_{s=3}^{N}J_{s}(z)W^{(s)}_{-s+1}\Bigr)dz\,,\quad \bar{a} = \Bigl(L_{-1} + \Tb(\zb)L_{1} +\sum_{s=3}^{N}\Jb_{s}(\zb)W^{(s)}_{s-1} \Bigr)d\zb\,,
\end{equation}
where the $\{L_0,L_{\pm1}\}$ generators correspond to the $sl(2)$ subalgebra and we have in addition $N-2$ multiplets $\{W_{m}^{(s)}\}$ with $s=2,\ldots,N$ and $m=-(s-1),\ldots, (s-1)$. The asymptotic symmetry algebra of the three-dimensional theory is then found to be $\mathcal{W}_{N}\oplus \mathcal{W}_{N}\,$, where $T(z)$ and $\Tb(\zb)$ transform as the the left- and right-moving components of the stress tensor, and $J_{s}(z)$, $\overline{J}_{s}(\zb)$ as primary operators of weights $(s,0)$ and $(0,s)$ \cite{Henneaux:2010xg,Campoleoni:2010zq}.

The task at hand consists of constructing the connections encoding the data dictated by the configuration in the CFT. For the purpose of computing R\'enyi entropy, we can think of the problem geometrically as the connection that supports the replica Riemann surface on the boundary via appropriate monodromy conditions on the bulk gauge fields. Putting back the radial dependence of the connection, it is easy to see that the currents $T(z)$, $J_{s}(z)$ (and similarly in the other chiral sector) correspond to the normalizable modes of the bulk fields. As usual in holographic dualities, they are then identified with the one-point function of the operators in the dual CFT. With the replica boundary conditions in place, the key entry of the holographic dictionary is then
\begin{equation}\label{dictionary 1}
T(z) = \bigl\langle \hat{T}\bigr \rangle_{\mathcal{R}_{n,N_{I}}}\,,\qquad J_{s}(z) =\bigl\langle \hat{J}_{s}\bigr \rangle_{\mathcal{R}_{n,N_{I}}}~,
\end{equation}
\noindent where $\hat{T}$ and $\hat{J}_{s}$ are the stress tensor and current operators in the chiral algebra of the dual CFT, on the Riemann surface $\mathcal{R}_{n,N_{I}}\,$. Since correlators on the branched cover $\mathcal{R}_{n,N_{I}}$ can be rewritten using the twist operators described in section \ref{sec:renyi}, we arrive at the alternative representation of the dictionary:
\begin{align}\label{dictionary 2}
T(z)
={}&
\frac{\bigl\langle \hat{T}^{(n)}(z) \twist(z_1,\overline{z}_1)\atwist(z_2,\overline{z}_2)\ldots \twist(z_{2N_I-1},\overline{z}_{2N_I-1})\atwist(z_{2N_I},\overline{z}_{2N_I}) \bigr\rangle}{\bigl\langle \twist(z_1,\overline{z}_1)\atwist(z_2,\overline{z}_2)\ldots \twist(z_{2N_I-1},\overline{z}_{2N_I-1})\atwist(z_{2N_I},\overline{z}_{2N_I}) \bigr\rangle}~,
\\
J_{s}(z)
={}&
\frac{\bigl\langle \hat{J}_{s}^{(n)}(z) \twist(z_1,\overline{z}_1)\atwist(z_2,\overline{z}_2)\ldots \twist(z_{2N_I-1},\overline{z}_{2N_I-1})\atwist(z_{2N_I},\overline{z}_{2N_I}) \bigr\rangle}{\bigl\langle \twist(z_1,\overline{z}_1)\atwist(z_2,\overline{z}_2)\ldots \twist(z_{2N_I-1},\overline{z}_{2N_I-1})\atwist(z_{2N_I},\overline{z}_{2N_I}) \bigr\rangle}~,
\end{align}
\noindent where $ \hat{T}^{(n)}$ and $ \hat{J}_{s}^{(n)}$ are the stress tensor and higher spin currents in the cyclic orbifold $\text{CFT}^{n}/\mathds{Z}_{n}\,$. The latter are simply the sum of the corresponding operators over all copies of the theory, and in particular invariant under the replica symmetry.

As discussed in section \ref{sec:renyi} the expectation values in the above formulae can be taken in the vacuum or in excited states. More broadly, we can consider insertions of generic operators\footnote{It is often convenient to think of branch points as the insertion of twist operators, and treat them in the same footing as other operator insertions.} and demand that the currents in the connection are compatible with the local and global properties of these insertions. A fully general discussion can be quite cumbersome, but we will implement the following simplifications:
\begin{enumerate}
\item Euclidean time is not periodic, and hence we will not impose smoothness around a thermal cycle. In the CFT side, we then consider theories that were originally defined on the plane or the cylinder, which will simplify our task of building $T(z)$ and $J_{s}(z)\,$. We note however that studying the problem on the torus seems doable; see e.g. \cite{Barrella:2013wja}.

\item As a consequence of the above, we will not include sources for the currents. We will instead describe configurations carrying fixed charges, which is most natural in Lorentzian signature. Setting the sources to zero implies the boundary conditions $a_{\bar{z}} = \bar{a}_{z}=0\,$, so the connection $a$ is holomorphic while $\bar a$ is anti-holomorphic. In other words, we do not deform the boundary conditions \eqref{DS connections}.

\item For the purpose of computing R\'enyi entropies, we will assume that the replica symmetry is preserved in the bulk.
\end{enumerate}

As we will discuss below, imposing a suitable set of monodromy conditions fixes the general form of the stress tensor and higher spin currents. In what follows we will describe general aspects of the monodromy conditions that encode the data of the CFT operators and the topology of the replica manifold.

%%%%%%%%%%%%%%%%%%%%%%%%%%%%%%%%%%%%%%%%%%%%%%%%%%%%%
\subsection{Differential equation}\label{subsec:ODE}
%%%%%%%%%%%%%%%%%%%%%%%%%%%%%%%%%%%%%%%%%%%%%%%%%%%%%

Let us focus on a single chiral sector for simplicity. In order to compute the monodromies of the Drinfeld-Sokolov connections \eqref{DS connections} it is useful to introduce an auxiliary ODE
\begin{equation}\label{auxiliary ODE}
\partial \Psi = a(z)\Psi\,.
\end{equation}
\noindent Here $\Psi$ is an $N$-dimensional vector whose $i$-th component has the form $D^{(i-1)}(T,J_{s})\psi(z)$, where $\psi(z)$ is a scalar and $D^{(j)}(T,J_{s})$ denotes a differential operator of order $j$ acting on $\psi(z)$, so that the matrix ODE reduces to a single $N$-th order differential equation. The algorithm for determining the form of $\Psi$ is straightforward. Start from (\ref{auxiliary ODE}) with the components of $\Psi$ being independent. Then successively solve the equations, starting with the lowest order equation and working upwards. This determines $N-1$ of the components in terms of the remaining one. For example, in the $sl(2)$ case one has
\begin{equation}\label{eq:ode2}
\Psi =
\left(
\begin{array}{c}
-\partial \psi(z) \\
\psi(z)
\end{array}
\right)
\qquad \Rightarrow \qquad \partial^{2}\psi(z) + T(z)\psi(z) =0\,.
\end{equation}

\noindent Similarly, in the $sl(3)$ case
\begin{gather}\label{eq:ode3}
\Psi =
\left(
\begin{array}{c}
\partial^{2}\psi(z) + 2T(z)\psi(z)\\
\partial \psi(z) \\
\psi(z)
\end{array}
\right)
\\
\vphantom{a}
\nonumber
\\
\Rightarrow \qquad
\partial^{3}\psi(z) + 4T(z)\partial\psi(z) +2\Bigl[\partial T(z) - 2W(z)\Bigr]\psi(z)=0\,.
\end{gather}

Note that \eqref{eq:ode2} and \eqref{eq:ode3} take the same form as the CFT equations \eqref{eq:T1} and \eqref{DE} encoding the decoupling of the light degenerate operator $\hat{\psi}\,$. In other words, the differential equation relevant for the computation of semiclassical conformal blocks via the monodromy method is already built into the Drinfeld-Sokolov connections \eqref{DS connections} in a very natural way. In principle one can do even more, because the bottom component of the field $\Psi$ in (\ref{auxiliary ODE})
is a ${\cal W}_N$ primary, and it will have null states of levels $1$ through $N$. To find these null states, we can use
the fact that ${\cal W}_N$ transformations and the ${\cal W}_N$ arise from the gauge transformations which preserve the Drinfeld-Sokolov
form of the gauge field $a(z)$, and $\Psi$ must obviously transform with the same gauge parameter. From this one can deduce
the form of the OPE of all the higher spin currents with each component of the vector $\Psi$ and from this infer the precise
form of all the null vectors. We will, however, not need the detailed form of these null vectors in what follows.

In the general case, the space of solutions of the auxiliary ODE is $N$-dimensional, so we can choose a basis of linearly independent solutions $\Psi^{(i)}$, $i=1,\ldots, N$, and collect them into a \emph{fundamental matrix}
\be\label{aa:3}
\Phi(z) \equiv \left(\begin{array}{ccc} \Psi^{(1)} & \cdots &\Psi^{(N)} \end{array}\right)~.
\ee
\noindent The linear independence of the $N$ solutions is then equivalent to the invertibility of $\Phi(z)\,$. Let us assume that the matrix components of $a(z)$, namely the currents, are meromorphic functions. If we follow $\Phi(z)$ around a closed loop $\gamma$ in the complex $z$-plane, the result $\Phi_\gamma(z)$ is in general not equal to $\Phi(z)$, but rather
\be
\label{eqn:monodef}
\Phi_\gamma(z) = \mathcal{P}\big(e^{\oint_\gamma a}\big) \Phi(z) \equiv \Phi(z) M_\gamma \, .
\ee
This defines the monodromy matrix $M_\gamma\,$, which measures the lack of analyticity of $\Phi(z)$. A rearrangement of \eqref{eqn:monodef} yields
\be\label{eq:ma}
M_\gamma = \Phi(z)^{-1} \mathcal{P}\big( e^{\oint_\gamma a}\big) \Phi(z)
\ee
\noindent emphasizing the relationship between the holonomy built out of the flat connection $a(z)$, and the monodromy matrix $M_\gamma$: they belong to the same conjugacy class. Naturally, the same considerations apply to the other chiral sector and the corresponding connection $\bar{a}\,$. Notice that we can always redefine $\Phi(z)\rightarrow \Phi(z) g$ with $g\in GL(N,\mathds{C})$, and that
this will have the effect of conjugating the monodromy $M_{\gamma}$ by $g\,$. When considering the monodromy around different closed
loops, we should always work with a fixed choice for $\Phi(z)$, so that the ambiguity in $\Phi(z)$ has the effect of conjugating
all monodromies simultaneously by the same constant $g\,$.

At this stage it is rather clear that the problem of determining the currents $T(z)$ and $J_s(z)$ in the Chern-Simons connections will mimic the discussion in the CFT. In particular, equations \eqref{eq:T1} and \eqref{DE} capture the holonomies of the bulk connection, making the agreement evident. In what follows we will phrase various conditions on the currents in terms of holonomies of $a(z)\,$.

Before proceeding, it is worth mentioning one issue that can cause confusion.  When we impose conditions on the holonomy, it will sometimes be understood that this is defined up to an element of the center of the gauge group. In the following we will have occasion to perform gauge transformations that are non-single valued by an element of the center, and these change the holonomy around a closed loop accordingly.  The gauge fields are in the adjoint representation, and so of course transform trivially under the center.   Ambiguities in the meaning of ``trivial holonomy" can be resolved by matching the holonomy to that of global AdS$_3\,$, which represents a smooth connection. 

%%%%%%%%%%%%%%%%%%%%%%%%%%%%%%%%%%%%%%%%%%%%%%%%%%%%%
\subsubsection{Monodromy around singular points}
%%%%%%%%%%%%%%%%%%%%%%%%%%%%%%%%%%%%%%%%%%%%%%%%%%%%%

Denote by $z_{i}$ a potential singularity in the connection, which could be a branch point, the position of a primary operator insertion, etc. We now consider the monodromy of the Drinfeld-Sokolov connection \eqref{DS connections on the plane} around $z_{i}\,$,
\begin{equation}
M_{i} \simeq \mathcal{P}e^{\oint_{C_{i}}a}\,,
\end{equation}
\noindent where the contour $C_{i}$ is a small loop enclosing $z_{i}\,$ and no other singularities, and $\simeq$ means that the constant monodromy matrix $M_{i}$ is in the same conjugacy class as the holonomy. %We will as before assume that the matrix components of the connection $a$ are meromorphic functions around $z=z_{i}\,$.

By performing gauge transformations on $a(z)$ one can reduce the order of the pole at $z_{i}$ to some minimal value dubbed the Poincar\'e rank $r_{P}$ (see e.g. \cite{Castro:2013kea,Castro:2013lba}). What this means is that there exists a gauge where $a_{z}(z)$ takes the form
\begin{equation}\label{Poincare rank}
a_{z}(z) \xrightarrow[]{z\to z_{i}}(z-z_{i})^{-r_{P}-1}a_{0}(z)\,,
\end{equation}
\noindent where $a_{0}(z)$ has a convergent Taylor series expansion around $z = z_{i}\,$, and $a_{0}(z_{i})$ is non-degenerate.  When $r_{P}=0$ the point $z_{i}$ is at most a regular singularity of the differential equation \eqref{auxiliary ODE}, associated with a pole in $a(z)$ and a branch cut in $\Phi(z)$. In particular, for loops enclosing a single pole, $r_{P} =0$ implies that the path ordering becomes trivial in the limit that the loop approaches the pole, and
\begin{equation}
r_{P}=0:\qquad M_{i}
% \simeq \exp \oint_{C_{i}}\frac{a_{0}(z)}{z-z_{i}}
\simeq e^{2\pi i a_{0}(z_{i})}
\end{equation}
\noindent in this case.  From the bulk perspective, we would like the gauge connections to have ``well-behaved" monodromy in this sense, and we will then require the singularities in the currents to have Poincar\'e rank zero.

Recall now the adjoint action of the $L_0$ generator: $e^{-x L_0}W^{(s)}_me^{xL_0}=e^{-m x}W_{m}^{(s)}\,$. Setting $x=\ln(z-z_{i})$ and acting on \eqref{DS connections}, the gauge-transformed connection reads\footnote{Note that this gauge transformation is only single valued up to an element of the center.}
\begin{align}
\tilde{a}_{z} ={}& e^{-\ln(z-z_{i})L_0}\left(a_{z} +\partial_{z}\right)e^{\ln(z-z_{i})L_0}
\\
%={}&
%\frac{L_1+L_0}{z-z_{i}} + \sum_{s = 2}^{N}\frac{\alpha_{s}}{k}Q_{s}(z)\left(z-z_{i}\right)^{s-1}W_{-(s-1)}^{(s)}
%\\
={}&
\left(z-z_{i}\right)^{-1}\left[L_1 +L_0+ \left(z-z_{i}\right)^{2}T(z)L_{-1} + \sum_{s = 3}^{N}J_{s}(z)\left(z-z_{i}\right)^{s}W_{-(s-1)}^{(s)} \right].
\end{align}
\noindent It follows that $a$ will have well-behaved monodromy around the $z_{i}$ provided
\begin{equation}\label{charge expansion}
T(z) \xrightarrow[]{z\to z_{i}} \frac{h_{i}}{\left(z-z_{i}\right)^{2}} + \ldots\,,\qquad J_{s}(z) \xrightarrow[]{z\to z_{i}} \frac{q^{(s)}_{i}}{\left(z-z_{i}\right)^{s}} + \ldots\,.
\end{equation}
\noindent Transforming to the cylinder via $z-z_{i}=e^{iw}$ one obtains the zero modes
\begin{equation}
T(w) = - h_{i} + \frac{1}{4} \,,\qquad J_{s}(w) = (-i)^{s}q^{(s)}_{i}\,,
\end{equation}
\noindent showing that the above connections describe the insertion of operators of conformal weight $h_{i}$ and charges $q_{i}^{(s)}$ (up to normalization). For this class of solutions the residue matrix is simply
\begin{equation}\label{residue matrix}
a_{0}(z_{i}) = L_1 +L_0+ h_{i} L_{-1}+ \sum_{s = 3}^{N}q_{i}^{(s)}W_{-s+1}^{(s)}
\end{equation}
\noindent and it has full rank if the $h_{i}\,$, $q_{i}^{(s)}$ are independent. Summarizing, the eigenvalues of the matrix \eqref{residue matrix} determine the conjugacy class of $M_{i}$ around the insertion at $z_{i}\,$.

It is worth emphasizing that in situations where eigenvalues of the residue matrix differ by an integer, such as e.g. $h_{i}=q_i^{(s)}=0\,$, extra care has to be exercised in computing the monodromy. In this case $M_i$ could have a non-trivial Jordan form, signaling that the associated ODE admits logarithmic branches of solutions on special slices in parameter space. We discuss such an example in appendix \ref{subsubsec: resonant}.

Let us now discuss the monodromy around $z = \infty\,$, which constrains subleading terms in the expansions \eqref{charge expansion}. When there are no operators inserted at $z = \infty\,$, one requires the connection to have trivial monodromy around infinity. This requirement is equivalent to the usual notion of smoothness of the currents:
\begin{equation}\label{vacuum conditions at infinity}
T(z\to \infty) \sim \frac{1}{z^{4}}\,,\qquad J_{s}(z\to\infty) \sim \frac{1}{z^{2s}}\,,
\end{equation}
\noindent which follows by e.g. using the coordinate $\zeta = 1/z$ and demanding finiteness as $\zeta \to 0\,$. If an operator of charges $(h_{\infty},q_{\infty}^{(s)})$ is inserted at infinity, we instead require\footnote{In a slight abuse of notation, we use $ T(\zeta)\equiv T_{\zeta\zeta}(\zeta) = z^{4}T(z)$ and $ J_{s}(\zeta) \equiv (-1)^{s}z^{2s}J_{s}(z)\,$.}
\begin{equation}\label{insertions at infinity}
T(\zeta \to 0) = \frac{h_{\infty}}{\zeta^{2}}+\ldots \,,\qquad J_{s}(\zeta \to 0) = \frac{q^{(s)}_{\infty}}{\zeta^{s}} + \ldots
\end{equation}
\noindent as in \eqref{charge expansion}.

\subsubsection{Example: $sl(2)$}\label{sec:cssl2}
In the $sl(2)$ case the residue matrix \eqref{residue matrix} reduces to
\begin{equation}
a_{0}(z_{i}) = L_1 +L_0 + h_{i} L_{-1} =
\left(\begin{array}{cc}
1/2 & h_{i} \\
-1& -1/2
\end{array} \right)
\end{equation}
\noindent and therefore
\begin{equation}\label{eq:mi2}
M_{i} \simeq -e^{2\pi i a_{0}} \simeq - \left(\begin{array}{cc}
e^{2\pi i\lambda} & 0 \\
0& e^{-2\pi i\lambda}
\end{array} \right)\,,\qquad \text{with}\quad \lambda = \frac{1}{2}\sqrt{1-4h_{i}}\,.
\end{equation}
\noindent

%Let us now specialize to the case where region $A$ is a single-interval and the theory is in the vacuum state.
Let us now specialize to the case where we have two insertions. We set the weights $h_{1} = h_{2} \equiv h$ of the insertions at the endpoints $z_{1}$, $z_{2}$ of the interval, and demand the connection to have trivial monodromy around infinity. Writing
\begin{equation}\label{one interval stress tensor}
T(z) = \frac{h}{(z-z_{1})^{2}} + \frac{h}{(z-z_{2})^{2}} + \frac{c_{1}}{z-z_{1}} + \frac{c_{2}}{z-z_{2}}
\end{equation}
\noindent from \eqref{vacuum conditions at infinity} we get
\begin{equation}\label{single interval accessory parameters}
c_{1} = -c_{2} = \frac{2h}{z_{2}-z_{1}}~. %= \frac{1-n^{-2}}{2(z_{2}-z_{1})}\,.
\end{equation}
\noindent Note that smoothness at infinity also precludes the appearance of additional analytic terms in \eqref{one interval stress tensor}. In the simple case with two insertions, the requirement of trivial monodromy at infinity is then enough to fix the \textit{accessory parameters} $(c_{1}, c_{2})$ in terms of the dimension of the operators. This is no longer the case for multiple insertions, as we will discuss below. Obviously, the result above also follows directly by considering the correlation function of $T(z)$ with
two primaries and using the Ward identity for $T(z)$.

\subsubsection{Example: $sl(3)$}\label{sec:sl3m}
As a second example, we consider the $sl(3)$ theory in a case with four insertions and charges assigned as
%%
%\bea
%z&=&0:\quad (h_2,q_2) \cr
%z&=&x:\quad (h_1,q_1) \cr
%z&=&1:\quad (h_1,-q_1) \cr
%z&=&\infty :\quad (h_2,-q_2)~.
%\eea
%%
\begin{alignat}{3}
z={}& 0:& &\qquad & &(h_2,q_2) \cr
z={}&x:& &\qquad & &(h_1,q_1) \cr
z={}&1:& &\qquad & &(h_1,-q_1) \cr
z={}&\infty :& &\qquad & & (h_2,-q_2)~.
\end{alignat}
\noindent This configuration includes as a particular case the single-interval cut in an excited state created by the operator of dimension $h_{2}\,$ and spin-3 charge $q_{2}\,$, in which case the insertions at $z=1$ and $z=x$ are branch points with $q_{1}=0$ and $h_{1} = (1/4)(n-1/n)$ (see below).

Based on the above discussion, the general expressions for the currents $T$ and $W$ consistent with the assumed singularities is
\bea\label{sl3TW}
T(z)&=& {h_2\over z^2} + {h_1 \over (z-x)^2} + {h_1 \over(z-1)^2} + \frac{c_{0}}{z} + \frac{c_{1}}{z-1} + \frac{c_{x}}{z-x}\\
W(z) &=& {q_2 \over z^3}+{q_1\over (z-x)^3}-{q_1\over (z-1)^3}+ {a_0\over z^2}+{b_0\over z}+ {a_x\over (z-x)^2}+{b_x\over z-x}+ {a_1\over (z-1)^2}+{b_1\over z-1}\,.\cr &&
\nonumber
\eea
\noindent Imposing the behavior \eqref{insertions at infinity} at infinity (with $h_{\infty}=h_{2}$ and $q^{(3)}_{\infty} = -q_2$) we obtain the constraints
\bea\label{acces}
c_0 &=& 2h_{1}-c_{x} + xc_{x}\cr
c_{1} &=& -2h_{1}-xc_{x} \cr
a_0& =& -a_x+{1\over 2}(b_0+b_x) +(a_x-b_x)x+{1\over 2} b_x x^2 \\
a_1&=& {1\over 2}(b_0 +b_x)-x a_x -{1\over 2} b_x x^2 \cr
b_1& =& -b_0-b_x~.\nonumber
\eea
\noindent These relations leave $c_{x},a_{x},b_{x}$ and $b_{0}$ undetermined. Depending on the nature of the problem, these parameters can be further constrained by imposing additional conditions around a closed path that encircles the points e.g. $z = x$ and $z = 1\,$. In section \ref{sec:pert} we will discuss how this condition can be implemented in practice.

\subsubsection{Branch cuts}\label{subsubsec: branch cuts}

Let us now comment on the case where the insertion $z_{i}$ is a branch point. Branch points are merely curvature singularities and we return back to the starting point after circling around them $n$ times. Hence, for a branch point one requires that the $n$-th power of the monodromy around $z_{i}\,$ is trivial (possibly up to an element of the center).  For standard entanglement entropy calculations, the latter monodromy condition amounts in practice to
\begin{equation}\label{branch point monodromy}
{\rm eigenvalues}\Bigl[\left(M_{i}\right)^{n} \Bigr]= \pm{\rm eigenvalues}\Bigl[{e^{2\pi i L_0}}\Bigr]\,,
\end{equation}
\noindent because $e^{2\pi i L_0} = \pm \mathds{1}$ is in the center of the gauge group. The choice of plus or minus is fixed by picking the element of the center that matches with the holonomy  along $\phi\sim\phi+2\pi$ of global AdS$_3\,$.

%This condition induces degenerate eigenvalues for $M_i$, and hence we can allow for non-trivial Jordan forms as the most general form of the monodromy matrix. This will allow for log solutions in the multi-%sheeted replica manifold.

Consider the example in \ref{sec:cssl2}: imposing \eqref{branch point monodromy} on \eqref{eq:mi2} gives
\begin{equation}\label{twist operator dimension}
%\left(M_{i}\right)^{n} = e^{2\pi i L_0} = -\mathds{1} \qquad \Rightarrow \qquad
n\lambda = \frac{1}{2} \qquad \Rightarrow \qquad h_{i} = \frac{1}{4}\left(1-\frac{1}{n^{2}}\right)\,.
\end{equation}
\noindent In other words, close to the branch points the stress tensor takes a form consistent with the insertion of an operator of dimension\footnote{The extra factor of $n$ in $\Delta$ below is due to the fact that the full stress tensor in the orbifold theory contains a sum over copies.}
\begin{equation}
\Delta = \frac{nc}{6}h_{i}=\frac{c}{24}\left(n-\frac{1}{n}\right),
\end{equation}
\noindent which is known to be the dimension of the twist operators enacting the replica symmetry \cite{Calabrese:2004eu}. It is worth emphasizing that this result holds for any $N >2$ as well, because the standard branch point twist operators do not carry higher spin charges, and the diagonalization of the residue matrix reduces to the $sl(2)$ block.

For calculations of the generalized entanglement entropy proposed in \cite{Hijano:2014sqa}, in which the corresponding branch point twist operators carry higher spin charges, we expect more generally
\begin{equation}
\text{eigenvalues}\Bigl[\left(M_{i}\right)^{n} \Bigr]=\text{eigenvalues}\,\biggl[\exp \Bigl(\alpha_{2} L_0 + \sum_{s=3}^{N} \alpha_{s}W_{0}^{(s)}\Bigr) \biggr]\, ,
\end{equation}
\noindent where the coefficients $\alpha_{s}$ are adjusted such that the group element on the r.h.s. belongs to the center of the gauge group. This more general condition can be interpreted as requiring that, after circling around the branch points $n$ times, we return back to the starting point up to a higher spin transformation that acts trivially on the higher spin fields, but possibly nontrivially on matter fields. Similar conditions have been imposed in supersymmetric R\'enyi entropies \cite{Crossley:2014oea,Alday:2014fsa,Hama:2014iea}. In appendix \ref{app:s3} we elaborate further on this generalized notion of entanglement.

%%%%%%%%%%%%%%%%%%%%%%%%%%%%%%%%%%%%%%%%%%%%%%%%%%%%%
\subsection{Variation of the action}
%%%%%%%%%%%%%%%%%%%%%%%%%%%%%%%%%%%%%%%%%%%%%%%%%%%%%
By now we have described how to use a suitable set of monodromy conditions that fix the expectation values of the stress tensor and higher spin currents, and consequently of the Drinfeld-Sokolov boundary connections. In order to obtain R\'enyi entropies, the remaining task is to evaluate the Chern-Simons action on this solution and obtain the saddle-point approximation to the partition function on the branched cover. In practice, evaluating the on-shell action requires a rather involved regularization procedure, but for present purposes this can be circumvented by computing instead the variation of the action with respect to the positions of the branch points, and integrating the resulting differential equations, along the lines of \cite{Hartman:2013mia,Faulkner:2013yia}.

As discussed above (c.f. \eqref{dictionary 1}-\eqref{dictionary 2}), the expression $T(z)$ that appears in the gauge connection obeys the properties of a CFT stress tensor. In particular, when we compute AdS correlation functions involving the stress tensor, $\langle \hat{T}(z) {\cal O}(z_1) \ldots \rangle$, these will be compatible with the operator product expansion
\be
\hat{T}(z) {\cal O}(z_1) \sim {h_{\cal O} \over (z-z_1)^2} + {6\over c}{1\over z-z_1} \p {\cal O}(z_1) +\ldots~.
\ee
Now, a correlation function of twist operators is equal to the bulk partition function with boundary conditions specified by the branched cover (along with boundary condition at past and future infinity corresponding to being in an excited state),
\be\label{Tcor}
\bigl \langle {\cal O}_2 \bigl | \twist (x) \atwist(1) \bigr |{\cal O}_2 \bigr \rangle = e^{-S_{\rm bulk}}~.
\ee
Furthermore, the expression for $T(z)$ written in (\ref{sl3TW}) is to be identified with the ratio of correlators with and without insertion of $\hat{T}(z)$ (c.f. \eqref{dictionary 2} applied to an excited state)
\be
T(z) = { \bigl \langle {\cal O}_2 \bigl | \hat{T}^{(n)}(z) \twist(x) \atwist(1) \bigr |{\cal O}_2 \bigr\rangle \over \bigl \langle {\cal O}_2\bigl |\twist (x) \atwist(1) \bigr|{\cal O}_2 \bigr\rangle}~.
\ee
\noindent Putting these facts together, we see that if $T(z) \sim {h_1 \over (z-x)^2} +{c_x \over z-x}+\ldots$ as $z\rt x$, then
\be\label{px}
c_x = -{6\over c} {\p S_{\rm bulk}\over \p x}~.
\ee
Given $c_x\,$, this equation is integrated to obtain $S_{\rm bulk}\,$, and then (\ref{Tcor}) gives the correlation function of interest.

To derive the same result more directly from Chern-Simons theory would involve performing a suitable diffeomorphism which
moves one of the points but will not change the action. The diffeomorphism will, however, change the metric (or rather
the complex structure) of the boundary. To undo this change, we need to perform a suitable subsequent gauge transformation which
does change the action and which then will give rise to (\ref{px}) as well.

%%%%%%%%%%%%%%%%%%%%%%%%%%%%%%%%%%%%%%%%%%%%%%%%%%%%%%%%%%%%%%%%%%%%%%%%%%%%%%%%
\section{Perturbative solution of the monodromy problem}\label{sec:pert}
%%%%%%%%%%%%%%%%%%%%%%%%%%%%%%%%%%%%%%%%%%%%%%%%%%%%%%%%%%%%%%%%%%%%%%%%%%%%%%%%

As discussed above, thinking either in terms of the vacuum block for a CFT with $\CW_N$ symmetry, or in terms of a Chern-Simons connection with prescribed boundary conditions, we are led to the same monodromy problem. In particular, we are instructed to consider an $N$-th order ODE on the complex $z$-plane. In this section we will solve this monodromy problem perturbatively in the case of four operator insertions: two heavy operators and two light operators. Here, a light operator is one whose rescaled charges are small, $h, q^{(s)} \ll 1$, so we can carry out perturbation theory in these quantities.

Recall that the insertion of an operator at $z=z_i$, in this language, corresponds to a regular singular point which creates a pole in $a(z)$ and a branch cut in $\Phi(z)$ at $z=z_i\,$. As mentioned above we are interested in four insertions, of two heavy operators (of the same species) and two light operators (of the same species). For concreteness, we assign a position and monodromy matrix to each insertion:
%%
%\bea
%z&=&0:\quad M_0 ~~ {\rm heavy}~,\cr
%z&=&x:\quad M_x ~~{\rm light}~, \cr
%z&=&1:\quad M_1 ~~ {\rm light}~,\cr
%z&=&\infty :\quad M_\infty ~~ {\rm heavy}~.
%\eea
%%
\begin{alignat}{3}
z={}&0:& \qquad &M_0&  & \,\text{ heavy}\,,
\nonumber\\
z={}&x:& \qquad  &M_x& &\,\text{ light}\,,
\nonumber\\
z={}&1:&\qquad  &M_1& &\,\text{ light}\,,
\\
z={}&\infty :& \quad &M_\infty& &\,\text{ heavy}\,.
\nonumber
\end{alignat}
The monodromy matrices are defined according to which singular point they enclose, however they depend not only on the local data at the singularity (i.e. charges of the operators) but as well on all coefficients in $a(z)$ since they are sensitive to the base point used for the contour $\gamma\,$.
However, for a regular singular point, the eigenvalues of $M_i$ depend only on the residue of $a(z)$ at the location of the operator insertion (local data). Thus different choices of contour yields monodromy matrices related by a similarity transformation.

We are interested in computing the vacuum block, which means that we should impose trivial monodromy around a contour that encloses $z=x$ and $z=1$, while not enclosing $z=0\,$. If we choose the contours defining $M_1$ and $M_x$ to share a base point then we would demand
\bea\label{eq:mtrivial}
M_1 M_x = \mathds{1} ~.
\eea
This condition is not automatic, and imposing it is the key of our analysis. A simple way to see that \eqref{eq:mtrivial} is non-trivial goes as follows.
If two operators are of the same kind the monodromy matrices are not necessarily equal, however their eigenvalues are related. For instance, since at $z=x$ and $z=1$ we have the same light operator; it must hold that
\be\label{eq:mu}
M_{1}= U^{-1} M_x^{-1} U~, \quad U\in GL(N,\mathds{C})~.
\ee
In other words their eigenvalues are related, and the inverse is due to the relative orientation of the insertion of each operator. $U$ is a matrix that brings the monodromies to a common basis and it depends generically on all coefficients of the ODE. Consistency between \eqref{eq:mtrivial} and \eqref{eq:mu} imposes restrictions on the components of $U$.

From the CFT perspective, imposing trivial monodromy picks out the vacuum block. Other blocks are obtained from nontrivial monodromy: replace the r.h.s. of \eqref{eq:mtrivial} such that $M_1M_x$ encodes the charges of the appropriate primary $\alpha\,$. For some CFTs, it is expected that the vacuum block is the dominant contribution to the four-point function in the large $c$ limit. To argue that the vacuum block dominates, one would need additional assumptions about the spectrum of light operators in the CFT \cite{Hartman:2013mia}, among perhaps other conditions.

From the Chern-Simons point of view trivial monodromy is the condition of vanishing holonomy for the connection, which in turn means that the cycle can be smoothly contracted in the bulk without encountering any nonzero field strength. This parallels the CFT side, since to compute a non-vacuum block we would expect to need additional matter in the bulk, and this matter would give rise to nonzero field strength.

To enforce \eqref{eq:mtrivial}, we will solve the ODE perturbatively. We choose as a small parameter the charges of the operators which we denoted light. Take
\be\label{eq:A0A1}
a(z)=a^{(0)} +\varepsilon a^{(1)}~,
\ee
where $\varepsilon$ controls the ``lightness'' of the operators at $z=1,x\,$. We split the fundamental matrix as $\Phi = \Phi_0 \Phi_1$ where
\be\label{eq:ap0}
(\partial -a^{(0)})\Phi_0 =0~,
\ee
and hence
\be
(\partial -\varepsilon \Phi_0 ^{-1} a^{(1)}\Phi_0 )\Phi_1=0~.
\ee
To linear order in $\varepsilon$, the solution is
\be\label{eq:P1}
\Phi_1= \mathds{1} + \varepsilon \int dz\, \Phi_0 ^{-1} a^{(1)}\Phi_0 + O( \varepsilon^2)~.
\ee
Imposing that $M_1M_x= \mathds{1}$, implies that for a loop enclosing $z=1,x$ we must have
\be\label{eq:M1x}
\int_{\gamma=\{1,x\}} dz\, \Phi_0 ^{-1} a^{(1)}\Phi_0 =0~.
\ee
This equation will fix certain coefficients in $a^{(1)}$ up to linear order in $\varepsilon$. In the examples below we will see explicitly how it constraints the accessory parameters introduced in the previous sections.

Independent of the expansion in $\varepsilon$, one could solve the monodromy condition perturbatively in $x$ (s-channel) or $1-x$ (t-channel) while keeping $\varepsilon$ fixed (i.e. without assuming that the operators are light).

%%%%%%%%%%%%%%%%%%%%%%%%%%%%%%%%%%%%%%%%%%%%%%%%%%%%%
\subsection{Example: $N=2$}
%%%%%%%%%%%%%%%%%%%%%%%%%%%%%%%%%%%%%%%%%%%%%%%%%%%%%

We now reproduce the result obtained in \cite{Fitzpatrick:2014vua}.
We are interested in the case of four operator insertions: two heavy operators, and two light operators. The stress tensor in this case reads
\be\label{eq:tt2}
T(z)={{h_2}\over{z^2}}+\left({{1}\over{(z-1)^2}}+{{1}\over{(z-x)^2}}+{{2}\over{(1-z)z}}\right)h_1-c_x{{x(1-x)}\over{z(1-z)(z-x)}}~,
\ee
where we assigned charges as
\bea
&z=0: &~ h_2~ {\rm heavy} ~,\cr
&z=x:&~ h_1 ~ {\rm light}~,\cr
&z=1:&~ h_1~ {\rm light} ~,\cr
&z=\infty :&~ h_2~ {\rm heavy}~.
\eea
Here will scale like $h_1\sim \varepsilon$ and $c_x\sim \varepsilon$, while $h_2$ is fixed. Equation \eqref{eq:tt2} can be obtained either from the CFT arguments in section \ref{sec:cftsl2} or from the regularity condition in the Chern-Simons theory discussed in section \ref{sec:cssl2}.

In the notation \eqref{eq:A0A1} the zeroth order piece (which is independent of $\varepsilon$) is
\be
a^{(0)}= \left(\begin{array}{cc}0 & T^{(0)}\\ -1 & 0 \end{array}\right) ~,\quad T^{(0)}\equiv {{h_2}\over{z^2}}~,
\ee
and the fundamental matrix associated to the zeroth order equation is
\be\label{eq:p2}
\Phi_0= \left(\begin{array}{cc}-\partial \psi_1^{(0)} & -\partial \psi_2^{(0)}\\ \psi_1^{(0)} & \psi_2^{(0)} \end{array}\right)~,
\ee
where $\psi_i^{(0)}$ are the solutions to $\partial^2_z\psi^{(0)}+T^{(0)}\psi^{(0)}=0 $ which gives
\be
\psi_1^{(0)}=z^{(1+\alpha)/2} ~,\quad \psi_2^{(0)}=z^{(1-\alpha)/2}~, \quad \alpha = \sqrt{1-4h_2}~.
\ee
The terms that scale with $\varepsilon$ are
\be\label{eq:ca1}
a^{(1)}= \left(\begin{array}{cc}0 & T^{(1)}\\ 0 & 0 \end{array}\right) ~,
\ee
with
\be
T^{(1)}=\left({{1}\over{(z-1)^2}}+{{1}\over{(z-x)^2}}+{{2}\over{(1-z)z}}\right)h_1-c_x {{x(1-x)}\over{z(1-z)(z-x)}}~.
\ee
We want to impose \eqref{eq:M1x}; using \eqref{eq:p2} and \eqref{eq:ca1} we find
\be
\Phi_0 ^{-1} a^{(1)}\Phi_0 =
\left(\begin{array}{cc} T^{(1)} \psi_1^{(0)}\psi_2^{(0)} & T^{(1)} (\psi_2^{(0)})^2
\\ - T^{(1)} (\psi_1^{(0)})^2 & - T^{(1)} \psi_1^{(0)}\psi_2^{(0)} \end{array}\right)~.
\ee
Imposing \eqref{eq:M1x}, which is a contour that only encloses the poles at $z=x$ and $z=1$, requires that the sum of residues around these poles vanish. These residues are
%
%\bea
%&& {\rm Res}_{z=x} (z T^{(1)})+{\rm Res}_{z=1} (z T^{(1)}) =0~,\cr
%&&{\rm Res}_{z=x} (z^{1+\alpha} {c\over 6}T^{(1)})+{\rm Res}_{z=1} (z^{1+\alpha} {c\over 6}T^{(1)}) = [(1+\alpha)x^\alpha -1+\alpha] h_1 +(x^\alpha -1)x c_x~,\cr
%&&{\rm Res}_{z=x} (z^{1+\alpha} {c\over 6}T^{(1)})+{\rm Res}_{z=1} (z^{1+\alpha} {c\over 6}T^{(1)}) = [(1-\alpha)x^{-\alpha} -1-\alpha] h_1 +(x^{-\alpha} -1)x c_x~.
%\eea
\begin{align}
{\rm Res}_{z=x} \left(z T^{(1)}\right)+{\rm Res}_{z=1} \left(z T^{(1)}\right)
={}&
0\,,\nonumber\\
{\rm Res}_{z=x} \left(z^{1+\alpha} T^{(1)}\right)+{\rm Res}_{z=1} \left(z^{1+\alpha} T^{(1)}\right)
={}&
\bigl[(1+\alpha)x^\alpha -1+\alpha\bigr] h_1 -\left(x^\alpha -1\right)x c_x \,,\\
{\rm Res}_{z=x} \left(z^{1-\alpha} T^{(1)}\right)+{\rm Res}_{z=1} \left(z^{1-\alpha} T^{(1)}\right)
={}&
\bigl[(1-\alpha)x^{-\alpha} -1-\alpha\bigr] h_1 -\left(x^{-\alpha} -1\right)x c_x\, , \nonumber
\end{align}

\noindent and demanding that they vanish yields
\be
c_x={(1+\alpha)x^\alpha -1+\alpha\over x(x^\alpha -1)}h_1~.
\ee
From (\ref{eq:tt2}) we see that $c_x$ is the residue of the simple pole at $z=x\,$. According to the discussion that led to (\ref{px}) we can therefore compute the bulk action by integration,
\be
S_{\rm bulk} = {c\over 6} \int\! c_x\, dx
=
{c\over 6} \left[ 2\ln \left(1-x^\alpha \over \alpha\right)+(1-\alpha)\ln x \right]h_1~.
\ee
The correlation function is therefore
\be \label{correlator1}
\bigl\langle {\cal O}_2 \bigl | {\cal O}_{1}(x) \tilde{{\cal O}}_{1}(1) \bigr |{\cal O}_2 \bigr\rangle = e^{-S_{\rm bulk}} = x^{-{c\over 6}h_1}\left( x^{-{\alpha\over 2}}-x^{\alpha\over 2}\over \alpha\right)^{-{c\over 3}h_1}
\ee
which is the result obtained in \cite{Fitzpatrick:2014vua}.

This represents the correlator on the $z$-plane. To interpret the result it is convenient to map it to the cylinder, $z=e^{iw}\,$. Taking into account the conformal transformation of the operator at $x$, we find
\be\label{probecor}
\bigl\langle {\cal O}_2 \bigl | {\cal O}_{1}(w) \tilde{{\cal O}}_{1}(0) \bigr |{\cal O}_2 \bigr\rangle = {C \over \left[ \sin \left(\alpha w\over 2\right)\right]^{{c\over 3}h_1} }~.
\ee
As noted in \cite{Fitzpatrick:2014vua}, this has a simple bulk interpretation. Consider the conical defect metric
\be\label{conmet}
ds^2 = {\alpha^2 \over \cos^2 \rho}\left({1\over \alpha^2}d\rho^2 -dt^2+\sin^2 \rho d\phi^2\right)~.
\ee
The metric corresponds to a state with conformal dimension $({c\over 6}h_2,{c\over 6}h_2)$ with $\alpha=\sqrt{1-4h_2}\,\,$. Introduce a probe particle of mass $m={c\over 3}h_1\,$. The two-point function of the operator dual to this particle is obtained in the geodesic approximation as $e^{-m L}$, where $L$ denotes the (regularized) geodesic length. Letting the geodesic pierce the boundary at $\phi=t=0$ and at $w=\phi+it$, we find agreement with (\ref{probecor}). The heavy operator creates the background geometry, and the light operator corresponds to a probe in this geometry.

Another interesting observation made in \cite{Fitzpatrick:2014vua} is that for $h_2 >{1\over 4}$ the metric (\ref{conmet}) is a BTZ black hole. It is intriguing to see that this arises as the semiclassical description of a heavy operator insertion. We discuss this further in section \ref{discussion}.

If we take $h_1 = {1\over 4}(n-{1\over n})$ and $n \to 1$, then the light operator corresponds to a twist operator. The above result for the twist correlator will reproduce the Ryu-Takayanagi formula for the entanglement entropy in the metric (\ref{conmet}). This was discussed in \cite{Asplund:2014coa,Caputa:2014eta}.

%%%%%%%%%%%%%%%%%%%%%%%%%%%%%%%%%%%%%%%%%%%%%%%%%%%%%
\subsection{Example: $N=3$}
%%%%%%%%%%%%%%%%%%%%%%%%%%%%%%%%%%%%%%%%%%%%%%%%%%%%%

In the $sl(3)$ case, the analysis in section \ref{sec:block3} and \ref{sec:sl3m} instructed us to study the third order ODE
\be\label{ODE}
\partial^3\psi(z) + 4 T(z) \partial\psi(z)+2\partial T(z)\psi(z)-4W(z) \psi(z) =0~.
\ee
Recall that the stress tensor $T(z)$ and the spin-3 current $W(z)$ are meromorphic functions with prescribed singularities at the locations of operator insertions. This yields
\bea
T(z)&=& {h_2\over z^2} + {h_1 \over (z-x)^2} + {h_1 \over(z-1)^2} - {2h_1 \over z(z-1)}-{x(1-x)\over z(z-1)(z-x)}c_x~, \\
W(z) &=& {q_2 \over z^3}+{q_1\over (z-x)^3}-{q_1\over (z-1)^3}+ {a_0\over z^2}+{b_0\over z}+ {a_x\over (z-x)^2}+{b_x\over z-x}+ {a_1\over (z-1)^2}+{b_1\over z-1}~,\nonumber
\eea
where the seven constants (``accessory parameters") $(a_0, a_x, a_1, b_0, b_x, b_1, c_x)$ are subject to three relations, which can be written as
\bea\label{acces}
a_0& =& -a_x+{1\over 2}(b_0+b_x) +(a_x-b_x)x+{1\over 2} b_x x^2 ~,\cr
a_1&=& {1\over 2}(b_0 +b_x)-x a_x -{1\over 2} b_x x^2 ~,\\
b_1& =& -b_0-b_x~. \nonumber
\eea
The remaining free parameters are fixed by monodromy conditions. Our main interest is to extract the value of $c_x$, since the relation $c_x = {6\over c} {\partial S_{\rm bulk}\over \partial x}$ can then be integrated to find $S_{\rm bulk}$ that appears in the semi-classical $\CW_3$ block.

To proceed, we assume $h_1, q_1 \sim O(\varepsilon)$, and work to first order in these quantities. No assumption is made regarding the magnitude of $(h_2,q_2)$. We consider a closed path that encircles the points $z=x$ and $z=1$, but does not encircle $z=0\,$.
%The three linearly independent solutions of (\ref{ODE}) return to themselves up to a linear transformation, which defines the monodromy matrix.
We demand trivial monodromy, which is to say that we will impose \eqref{eq:mtrivial}.

%The problem is now sharply posed, but difficult to solve due to the challenge in solving the differential equation (\ref{ODE}).

With this in mind, we implement the perturbative expansion by writing
\be
a^{(0)}= \left(\begin{array}{ccc}0&-2T^{(0)}&4W^{(0)} \\ 1&0&-2T^{(0)} \\ 0&1&0\end{array}\right) ~,\quad
a^{(1)}= \left(\begin{array}{ccc}0&-2T^{(1)}&4W^{(1)} \\ 0&0&-2T^{(1)} \\ 0&0&0\end{array}\right)~,
\ee
with
\bea
T^{(0)}& = & {h_2\over z^2}~, \quad W^{(0)} = {q_2\over z^3}~, \cr
%&&\cr
T^{(1)}& = & {h_1 \over (z-x)^2} + {h_1 \over(z-1)^2} - {2h_1 \over z(z-1)}-{x(1-x)\over z(z-1)(z-x)}c_x ~,\cr
W^{(1)}& = &{q_1\over (z-x)^3}-{q_1\over (z-1)^3}+ {a_0\over z^2}+{b_0\over z}+ {a_x\over (z-x)^2}+{b_x\over z-x}+ {a_1\over (z-1)^2}+{b_1\over z-1}~.
\eea

We need to evaluate \eqref{eq:M1x}, and for that we need to build $\Phi_0$ as defined in \eqref{eq:ap0}. The zeroth order equation is
\be
{\p^3 \psi^{(0)}} +{4h_2 \over z^2}\p {\psi^{(0)}} -{4h_2 \over z^3}{\psi^{(0)}} -{4q_2\over z^3}{\psi^{(0)}}=0~,
\ee
and the three independent solutions are
\be \label{eq:p30}
\psi^{(0)}_{n} = z^{1+ p_n}~,\quad n =1,2,3~,
\ee
where $p_n$ are the three roots of
\be\label{cubic}
p^3 -(1-4h_2)p-4q_2 =0~.
\ee
Recall that these obey $p_1+p_2+p_3=0\,$. Using \eqref{aa:3}, \eqref{eq:ode3} and \eqref{eq:p30} gives $\Phi_0\,$. The combination of interest is
\be
a^{(1)}\Phi_0 = \left(\begin{array}{ccc}-2T^{(1)} \p \psi_1^{(0)} +4W^{(1)}\psi_1^{(0)}&-2T^{(1)} \p \psi_2^{(0)} +4W^{(1)}\psi_2^{(0)}&-2T^{(1)} \p \psi_3^{(0)} +4W^{(1)}\psi_3^{(0)} \\ -2T^{(1)}\psi_1^{(0)} & -2T^{(1)}\psi_2^{(0)}&-2T^{(1)}\psi_3^{(0)} \\ 0&0&0\end{array}\right)
\ee
and hence the relevant integrals we need to compute are
\begin{align}
M^{(1)}_{nm}
\equiv{}&
\oint_{\gamma} dz (\Phi_0^{-1}a^{(1)}\Phi_0)_{nm}
\\
={}&
{2(p_{n+1}-p_{m+2})\over \det \Phi_0} \oint_{\gamma} dz \le[ (p_n+p_m)z^{1+p_n-p_m} T^{(1)} -2z^{2+p_n-p_m}W^{(1)} \ri].
\nonumber
\end{align}

In order to extract the vacuum block we now need to impose the trivial monodromy condition $M^{(1)}_{nm}=0\,$. The diagonal equations $M^{(1)}_{nn}=0$ are easily seen to be equivalent to the equations (\ref{acces}). This leaves six equations for four free parameters; however, it turns out that only four of the equations are independent, leading to a unique solution. After a considerable amount of computer aided algebra, we obtain
%
%\bea\label{csol}
%&&\!\!\!\!\!\!\!\! \!\!\!\!\!\! c_x = \cr
%&&\!\!\!\! \!\!\!\! \!\!\!\! \!\!\!\! {\sum_n \Big[(p_{n+1}-p_{n-1})(p_n-p_{n+1})\big[ x^{p_n-p_{n-1}}(2+p_n-p_{n-1})+x^{-p_n+p_{n+1}}(2-p_n+p_{n+1})\big]+4p_n^2-4p_{n-1}p_{n+1}\Big] \over 2x \sum_n[x^{-p_n}(-p_{n-1}+p_{n+1})] \sum_n[x^{p_n}(p_{n-1}-p_{n+1})] }h_1 \cr && \cr
%&&\!\!\!\! \!\!\!\! \!\!\!\! \!\!\!\! +{\sum_n \Big[ (p_{n+1}-p_{n-1})(p_{n-1}-p_{n})(p_{n+1}-2p_{n-1}+p_n)(x^{p_{n+1}-p_n}+x^{p_{n}-p_{n+1}})\Big] +\prod_n(p_{n-1}-2p_n+p_{n+1})\over 2x \sum_n[x^{-p_n}(-p_{n-1}+p_{n+1})] \sum_n[x^{p_n}(p_{n-1}-p_{n+1})] }q_1 \cr &&
%\eea
%
\begin{equation}\label{csol}
c_{x} = \frac{C_{h} \, h_{1} + C_{q}\, q_{1}}{2x \sum_n[x^{-p_n}(-p_{n-1}+p_{n+1})] \sum_n[x^{p_n}(p_{n-1}-p_{n+1})]}
\end{equation}

\noindent where we used the shorthand
\begin{align}
C_{h}
={}&
\sum_n \Big[(p_{n+1}-p_{n-1})(p_n-p_{n+1})\big[ x^{p_n-p_{n-1}}(2+p_n-p_{n-1})+x^{-p_n+p_{n+1}}(2-p_n+p_{n+1})\big]
\nonumber\\
&\hphantom{\sum_n \Big[}
+4p_n^2-4p_{n-1}p_{n+1}\Big]
\\
C_{q}
={}&
\sum_n \Big[ (p_{n+1}-p_{n-1})(p_{n-1}-p_{n})(p_{n+1}-2p_{n-1}+p_n)(x^{p_{n+1}-p_n}+x^{p_{n}-p_{n+1}})\Big]
\nonumber\\
&
+\prod_n\left(p_{n-1}-2p_n+p_{n+1}\right).
\end{align}

To arrive at this form of the solution we used the relation $\sum_n p_n=0\,$. Given this result for $c_x\,$, it is not easy to evaluate $S_{\rm bulk}={c\over 6}\int \!c_x\, dx\,$. Fortunately, the corresponding bulk computation will yield $S_{\rm bulk}$ directly, and then we can confirm that it yields the same $c_x$ upon differentiation.

%%%%%%%%%%%%%%%%%%%%%%%%%%%%%%%%%%%%%%%%%%%%%%%%%%%%%%%%%%%%%%%%%%%%%%%%%%%%%%%%
\section{Wilson line computation of vacuum block }\label{wilson}
%%%%%%%%%%%%%%%%%%%%%%%%%%%%%%%%%%%%%%%%%%%%%%%%%%%%%%%%%%%%%%%%%%%%%%%%%%%%%%%%

In this section we evaluate the action for a Wilson line probe in an asymptotically AdS$_3$ background. Both the Wilson line and the background solution carry arbitrary spin-2 and spin-3 charges. We will establish, by direct computation, that this action computes the vacuum ${\cal W}_3$ block with two heavy operators (corresponding to the background) and two light operators (corresponding to the probe). In particular we will demonstrate that the result matches the result we obtained from the CFT/Chern-Simons monodromy computation (\ref{csol}). The Wilson line approach turns out to be a good deal more efficient, as it directly produces the vacuum block, bypassing the need to perform a final integration as is the case in the monodromy approach.

Special cases of this computation are relevant to entanglement entropy. In \cite{Ammon:2013hba}, and its equivalent formulation \cite{deBoer:2013vca}, a specific charge assignment for the probe, with vanishing spin-3 charge, was proposed to yield entanglement entropy. The present analysis puts this proposal on a firmer footing, since it demonstrates that the probe yields the vacuum block contribution to the correlation function of twist operators in the presence of other operators that set up an excited state. The missing step to prove that this probe computes entanglement entropy is to establish that only the vacuum block contributes in the limit of large central charge, a result which will require some additional assumptions about the spectrum of operators in the CFT, and which we do not delve into here. See \cite{Hartman:2013mia} for discussion of the necessary conditions in the case of Virasoro blocks.

A probe carrying nonzero spin-3 charge was argued in \cite{Hijano:2014sqa} to compute a generalized spin-3 version of entanglement entropy, an object that appears quite natural to define on the bulk side of the AdS/CFT correspondence, but whose meaning is at present obscure in the CFT. In \cite{Hijano:2014sqa} it was suggested that this spin-3 entropy could be computed in the CFT from the correlators of some sort of twist operators carrying nonzero spin-3 charge. Our present results will not really shed any new light as to the definition of these novel twist operators, but we will verify that the Wilson line probe can be used to compute the vacuum block contribution to the correlation function of these operators in an excited state.

We consider the following connection, corresponding to an asymptotically AdS$_3$ solution with cylindrical boundary, $w\cong w+2\pi$:
\bea\label{WLconn}
a &=& \Bigl(L_1 +T(w) L_{-1}+W(w) W_{-2}\Bigr)dw \\
\ab & =& L_{-1} d\wb~,
\eea
with $T(w)$ and $W(w)$ both constant, the dependence on $w$ being displayed just to remind us that these quantities are defined on the cylinder. We write
\be
T(w) = -h_2+{1\over 4}~,\quad W(w) = -iq_2~,
\ee
where $(h_2,q_2)$ are the charges carried by the operator that creates the excited state. In (\ref{WLconn}) we have chosen to turn on only holomorphic currents to reduce clutter, but it is straightforward to include their anti-holomorphic counterparts.

The general framework for defining and computing probe actions has been described in \cite{Ammon:2013hba}, and the specific computation done here is essentially the same as one appearing in \cite{Castro:2014mza}. We therefore just sketch the main steps.

The Wilson line is taken to extend between two points on the boundary, one of which we fix to be $w=0$ with the other left arbitrary. A large $\rho $ cutoff at $\rho=-\ln \eps$ is imposed to regulate divergences; this maps to a UV cutoff in the CFT according to the usual IR/UV relation in AdS/CFT.

We first define
\be
L= e^{-\rho L_0} e^{-a_w w}~,\quad R= e^{L_{-1}\wb}e^{-\rho L_0}~.
\ee
$L$ and $R$ are the gauge transformations that generate the flat connections (\ref{WLconn}) starting from ``nothing". The probe action is defined in terms of the matrix $M$, defined as\footnote{Here $\cong$ means conjugate to.}
\be
M = [R(s_i)L(s_i)][R(s_f)L(s_f)]^{-1} \cong e^{\ln \eps L_0} e^{a_w w} e^{-\ln \eps L_0}e^{-\ln \eps L_{-1}\wb}~.
\ee
As $\eps \rt 0$, the traces of $M$ behave as
\be
\Tr \left[M\right] = {m_1 \over \eps^4}+\ldots~,\quad (\Tr \left[M\right])^2-\Tr\left[M^2\right] = {2m_2 \over \eps^4}+\ldots
\ee
which defines the quantities $m_{1,2}\,$. In particular, the eigenvalues of $M$ as $\eps\rt 0$ behave as $\lambda_M \approx ({m_1 \over \eps^4},{m_2\over m_1}, {\eps^4 \over m_2})\,$. The probe action is expressed in terms of these eigenvalues as
\be
I = \Tr \Bigl[ \ln(\lambda_M) \Pc_0 \Bigr]
\ee
where for a probe carrying charges $(h_1,q_1)$ we have
\be
{6\over c} \Pc_0 = {h_1\over 2}L_0 + {3q_1\over 2}W_0~.
\ee
In our standard $sl(3)$ conventions with
\be
L_0= {\rm diag}(1,0,-1)~,\quad W_0 = {\rm diag} ({1\over 3}, -{2\over 3}, {1\over 3})
\ee
we thus have ${6\over c}\Pc_0 ={\rm diag}( {h_1+q_1\over 2}, -q , {-h_1+q_1\over 2})$ and
\be\label{probeI}
{6\over c}I = {h_1\over 2} \ln \left( m_1 m_2 \over \eps^8\right) +{3q_1\over 2} \ln \left( m_1 \over m_2\right)~.
\ee

It is not difficult to evaluate $m_{1,2}$, and we find
\bea
m_1 & =& {2\wb^2 \over \det_{mn} (p_m^{n-1})} \sum_{n=1}^3 (p_n-p_{n+1})e^{ip_{n+2}w}~, \cr
m_2& =& {2\wb^2 \over \det_{mn} (p_m^{n-1})} \sum_{n=1}^3 (p_n-p_{n+1})e^{-ip_{n+2}w}~,
\eea
where $p_n$ are the eigenvalues of $ia_w$ as given in \eqref{WLconn}. Here $p_{n+3} \equiv p_n\,$, and they satisfy the cubic equation
\be
p_n^3-(1-4h_2)p_n -4q_2 =0~,
\ee
which we recognize as being the same equation that appeared in (\ref{cubic}).
The probe action is read off from (\ref{probeI}).

The four-point function on the cylinder is then $e^{-I}$. To compare this to our previous computation we bring this to the $z$-plane via $z=e^{iw}\,$. Taking into account the conformal transformation of the operator at $w$, the four-point function on the plane is
\be
z^{-2h_1} e^{-I}\big|_{w=-i\ln z}= e^{ -\tilde{I}}~,\quad {6\over c}\tilde{I} = \left[ {h_1\over 2} \ln \left(z^2 m_1 m_2 \over \eps^8\right) +{3q_1\over 2} \ln \left( m_1 \over m_2\right)\right]\Bigg|_{w=-i\ln z}~.
\ee
The comparison with the monodromy-based result is obtained by writing $z=x$ and taking the $x$-derivative,
\bea
{6\over c} \p_x \tilde{I} &= &{1\over 2x}\biggl[2+{\sum_{n=1}^3 (p_n-p_{n+1})p_{n+2} x^{-p_{n+2}} \over \sum_{n=1}^3 (p_n-p_{n+1}) x^{-p_{n+2}} }- {\sum_{n=1}^3 (p_n-p_{n+1})p_{n+2} x^{-p_{n+2}} \over \sum_{n=1}^3 (p_i-p_{n+1}) x^{-p_{n+2}} } \biggr] h_1\nonumber\\
\vphantom{a}
\nonumber\\
&&\quad +{3\over 2x} \biggl[{\sum_{n=1}^3 (p_n-p_{n+1})p_{n+2} x^{p_{n+2}} \over \sum_{n=1}^3 (p_n-p_{n+1}) x^{p_{n+2}} }+ {\sum_{n=1}^3 (p_n-p_{n+1})p_{n+2} x^{-p_{n+2}} \over \sum_{n=1}^3 (p_n-p_{n+1}) x^{-p_{n+2}} } \biggr]q_1
\eea
We now compare this result to (\ref{csol}). As written the formulas appear different, but using $\sum_n p_n=0$ one can in fact show that $c_x = {6\over c}\p_x \tilde{I}\,$. We have therefore confirmed that the Wilson line computes the ${\cal W}_3$ vacuum block in the semi-classical limit, to linear order in the light charges.

%%%%%%%%%%%%%%%%%%%%%%%%%%%%%%%%%%%%%%%%%%%%%%%%%%%%%%%%%%%%%%%%%%%%%%%%%%%%%%%%
\section{The Toda perspective}\label{toda}
%%%%%%%%%%%%%%%%%%%%%%%%%%%%%%%%%%%%%%%%%%%%%%%%%%%%%%%%%%%%%%%%%%%%%%%%%%%%%%%%

Having discussed computations based on monodromies and Chern-Simons theory, and on Wilson lines, we now turn to yet another
perspective on the same type of computations, namely that of Toda theory. Recall that Liouville theory describes a theory of
$2d$ quantum gravity and encodes all universal correlation functions of the stress tensor in any conformal field theory. It can
be obtained by coupling a CFT to a background metric in conformal gauge, and then integrating out the degrees of freedom of
the CFT. It can also be obtained from $3d$ gravity with a negative cosmological constant by computing the partition function
with fixed boundary metric in conformal gauge.

Similarly, Toda theory is believed to arise as the effective action for CFTs coupled to higher spin background fields
in conformal gauge. Since a complete metric-like formulation of higher spin theories is unknown, it is difficult to verify
this directly. There are however several indirect arguments to support this statement, see e.g. \cite{deBoer:1991jc,deBoer:1993th}.

Toda theory for systems with ${\cal W}_N$
symmetry is a theory of $N-1$ scalar fields with background charge and with a potential term which is a sum of exponentials,
one for each simple root of $sl(N)$. Standard vertex operators for the scalar fields correspond to primaries of the
underlying $\mathcal{W}_N$ symmetry of Toda theory, and operators with arbitrary higher spin charges can be obtained in this way.
{The type of computation we have been doing corresponds to a four-point correlation function in Toda theory of two
``heavy'' and two ``light'' operators, where one works to first order in the quantum numbers of the light operators,
so that their backreaction can be neglected.} We will now first review such types of computations in Toda theory,
then revisit the relation between
Chern-Simons theory and Toda theory. Toda theory could be a
natural framework to connect the monodromy computation in section \ref{sec:pert} and the bulk Wilson line in section \ref{wilson}.

%%%%%%%%%%%%%%%%%%%%%%%%%%%%%%%%%%%%%%%%%%%%%%%%%%%%%
\subsection{Semiclassical correlators in Toda theory}
%%%%%%%%%%%%%%%%%%%%%%%%%%%%%%%%%%%%%%%%%%%%%%%%%%%%%
Toda theory for $SL(N)$ is a theory where the basic variable is a diagonal $SL(N)$ matrix $G_0(z,\bar{z})$. We can of course
parametrize $G_0$ with exponentials of scalar fields, but find it more convenient to work with $G_0\,$. The action of Toda theory
reads
\be
\label{todaaction}
S_{\rm Toda} = \kappa\int d^2 z\, \left( \frac{1}{2} {\rm Tr}\left[G_0^{-1}\partial G_0 G_0^{-1} \bar{\partial}G_0\right] - {\rm Tr}\left[G_0 L_{-1}
G_0^{-1} L_1\right] \right)~,
\ee
where the same $sl(2)$ generators $L_{\pm 1}$ are used as those which appear in the boundary conditions of Chern-Simons theory through
the Drinfeld-Sokolov connections (\ref{DS connections}), and $\kappa$ is some normalization constant.

Correlation functions are of the form
\be
\bigl\langle V_1(z_1)\ldots V_k(z_k)\bigr\rangle = \int {\cal D}G_0\, e^{-S_{\rm Toda}} \prod_i e^{{\rm Tr}\left[q_i \log G_0(z_i)\right]}~,
\ee
with $q_i$ an algebra-valued matrix that contains the information of the charges carried by $V_i(z)$.  The semiclassical answer is found by evaluating the integrand on the solution of the field equations
\be \label{toda eom}
\bar{\partial} \left(\partial G_0 G_0^{-1}\right) + \bigl[G_0 L_{-1} G_0^{-1},L_{1}\bigr] = \sum_i \frac{q_i}{\kappa} \delta^{(2)}(z-z_i)~.
\ee
Since $\log G_0$ will diverge logarithmically near $z=z_i\,$ this answer is in general divergent, but one can regulate
the theory by cutting out small discs around the points $z_i$ as explained for Liouville theory in \cite{Zamolodchikov:1995aa} and used
for Toda theory in e.g. \cite{Fateev:2007ab}. It is in general not possible to solve (\ref{toda eom}) exactly. However, when
we can separate the set of operators in a set of ``heavy'' and ``light'' operators\footnote{To make the identification of
heavy and light more explicit, one usually introduces a dimensionless coupling constant $b$, and heavy and light fields
are those for which the $q_i$ scale as $b^{-1}$ and $b$ respectively. In terms of conformal dimensions, heavy operators
have dimensions which scale as $c$ and light operators have dimensions which are of order unity as $b\rightarrow 0$ and
$c\rightarrow \infty\,$.   Note that this definition of ``light" differs from that in \ref{eq:lh}; however the difference is immaterial provided we work to first order in the light operator dimension.  In either case, we proceed by evaluating the light operators on the saddle point fixed by the heavy operators, and so the result is the same.  } we can proceed as follows. We first find the saddle point for the correlation function involving
only the heavy operators, $\hat{G}_0$, write $G_0=\hat{G}_0(1+\epsilon)$, solve (\ref{toda eom}) to first
order in $\epsilon$, and compute the correction to the saddle point to first order in $\epsilon$ as well.

By varying (\ref{toda eom}), we find that $\epsilon$ obeys the field equation
\be \label{lin eom}
\bar{\partial}\partial \epsilon + \Bigl[\hat{G}_0 \bigl[\epsilon,L_{-1}\bigr]\hat{G}_0^{-1},L_1\Bigr] =
\sum_i{}' q_i \delta^{(2)}(z-z_i)~,
\ee
where the sum on the right hand side only involves the light fields. Near $z=z_i\,$, we have
$\epsilon \sim \frac{q_i}{2\pi\kappa} \log |z-z_i|^2 + \ldots\,$. Now naively, if we perturb a saddle point to first order
the value of the on-shell action does not change since the original saddle point obeys the equation of motion. We
have to be careful here, because $\epsilon$ is divergent, but for the regularized theory the same statement remains true.
The only additional contribution to the semiclassical correlation function is coming from the light vertex operators
evaluated on the saddle point $\hat{G}_0\,$. Therefore, we obtain
\be \label{todaresult}
e^{-S_{\rm Toda}} = \bigl\langle V_1(z_1)\ldots V_k(z_k)\bigr\rangle_{\rm semiclassical}
  \sim e^{-S_{\rm heavy}} \prod_i{}' e^{{\rm Tr}\left[q_i \log \hat{G}_0(z_i)\right]}~,
\ee
where the product involves only the light operators, and $S_{\rm Toda}$ is the regulated semiclassical action. Our discussion has been somewhat sketchy, for example in
case there is a continuous family of saddle point solutions for the correlation function of heavy operators, one
is left with a finite dimensional integral on the right hand side.  For more discussion on these types of computations for
Liouville and Toda theory, see e.g. \cite{Fateev:2007ab,Harlow:2011ny}. We will elaborate on these finite dimensional integrals below,
as they will turn out to be crucial for our discussion.

%%%%%%%%%%%%%%%%%%%%%%%%%%%%%%%%%%%%%%%%%%%%%%%%%%%%%
\subsection{Chern-Simons theory versus Toda theory}
%%%%%%%%%%%%%%%%%%%%%%%%%%%%%%%%%%%%%%%%%%%%%%%%%%%%%
The boundary conditions for Chern-Simons theory involved the Drinfeld-Sokolov gauge fields (\ref{DS connections})
which we will rewrite as pure gauge as
\be \label{d0}
\partial g g^{-1} =  a= W+ L_1~,
\quad
\bar{g}^{-1} \bar{\partial} \bar{g} = \bar{a} =  \overline{W} + L_{-1}~,
\ee
where $W$ and $\overline{W}$ are holomorphic and anti-holomorphic and contain all the higher spin currents.
The group element or fundamental matrix $g(z)$ is in general multi-valued and it is the monodromy of $g(z)$
that we used to determine the contribution of the identity block to the four-point function, and the functions
appearing in the bottow row of $g(z)$ obey suitable $N$th-order differential equations.

Next, following \cite{Balog:1990mu} form the combination
\be \label{d1}
G=g(z)\bar{g}(\bar{z}) = G_- G_0 G_+~,
\ee
where the decomposition on the right hand side is in terms of matrices that have negative, zero and positive
grade with respect to the $sl(2)$ grading (in other words, $G_-$ is upper triangular and $G_+$ is lower triangular,  each with ones along the diagonal,  and $G_0$ is diagonal).
One can show, using the grading defined by the $sl(2)$ embedding, that (\ref{d0}) implies
\be \label{d2}
\partial G_+ G_+^{-1} = G_0^{-1} L_1 G_0~,
\quad
G_-^{-1} \bar{\partial} G_{-}  = G_0 L_{-1} G_0^{-1}~.
\ee

We know from (\ref{d1}) that $\bar{\partial} (\partial G G^{-1})=0\,$. In the following, we will show that this also happens
to be the usual equation of motion of WZW theory.
Inserting the decomposition (\ref{d1}) in this equation and using (\ref{d2}) we get
\be
\bar{\partial} \bigl(\partial G_- G_-^{-1} + G_- \partial G_0 G_0^{-1} G_-^{-1} + G_- L_1 G_-^{-1}\bigr)=0~.
\ee
The degree zero (i.e. diagonal) part of this equation, combined once more with (\ref{d2}), gives
\be
\bar{\partial} (\partial G_0 G_0^{-1}) + \bigl[G_0 L_{-1} G_0^{-1},L_1\bigr]=0\,,
\ee
which is precisely the classical Toda field equation (\ref{toda eom}).
Therefore there is a general way to construct classical solutions of Toda theory starting from a set of (anti-)holomorphic
higher spin currents.

If we plug (\ref{d1}) into (\ref{d0}), and use (\ref{d2}), the remaining equations are
\bea \label{d3}
\partial G_- G_-^{-1} + G_-(\partial G_0 G_0^{-1} + L_1)G_-^{-1} & = & L_1 + W ~,\nonumber \\
G_+^{-1}\bar{\partial} G_+ + G_+^{-1} (G_0^{-1}\bar{\partial} G_0 + L_{-1})G_+ & = &
L_{-1} +\overline{W}~,
\eea
which determine $G_-$ and $G_+$ exactly in terms of $G_0$ (i.e. there are no integration constants). Moreover,
this provides explicit expressions for $W$ and $\overline{W}$ in terms of $G_0\,$, which are precisely the expressions
for the conserved higher spin currents of Toda theory. This procedure is also known as the Miura transformation.

The above shows that the boundary conditions of Chern-Simons theory determine a solution of the Toda field equations, and conversely
a solution of the Toda field equations yields a suitable pair of gauge fields in Drinfeld-Sokolov form. This suggests that
one should be able to reformulate higher spin theories in $2+1$ dimensions, in such a way that the connection to Toda theory
becomes much more apparent, and it would be interesting to work this out in more detail. From a Chern-Simons point
of view, this probably would require us to work in a different gauge. As we mentioned before,
for ordinary gravity this corresponds to the case where the boundary metric is in conformal gauge, and one can easily work
out the corresponding gauge choice for Chern-Simons theory.

An important subtlety is that in our discussion the holomorphic and anti-holomorphic higher spin currents need not
be each other's complex conjugate. In particular, we can describe operators with different left and right
conformal dimensions and higher spin charges. Such operators do not exist in standard Toda theory, where the scalar
fields are real, and in order to accommodate such operators one must consider complexified solutions of Toda theory.

From (\ref{d3}), and with a bit of algebra, we deduce that the stress tensor that appears in the Drinfeld-Sokolov connection
is equal to
\be
T(z) = \frac{1}{{\rm Tr}\left[L_1L_{-1}\right]} \left( {\rm Tr}\Bigl[\left(G_0^{-1}\partial G_0\right)^{2}\Bigr] - \partial {\rm Tr} \left[L_0 G_0^{-1}\partial G_0\right] \right)~,
\ee
and this is, up to overall normalization, also the stress tensor of the Toda theory. We can evaluate this stress tensor
for the saddle-point solution of Toda theory which describes the correlation function of a combination of heavy and light
operators, again to first order in the light operators. Writing $G_0=\hat{G}_0(1+\epsilon)$, where
$\epsilon$ obeys the linearized field equation (\ref{lin eom}), we obtain
\be
T(z) = T^{\rm heavy}(z) +
\frac{1}{{\rm Tr}\left[L_1L_{-1}\right]} \left( 2{\rm Tr}\bigl[\partial \epsilon \hat{G}_0^{-1}\partial \hat{G}_0\bigr] -
{\rm Tr}\left[L_0 \partial^2 \epsilon\right] \right) + \ldots
\ee
and using the asymptotic behavior of $\epsilon$ near the insertion of a light operator, $\epsilon \sim \frac{q_i}{2\pi\kappa}
\log|z-z_i|^2$, we find that the expansion of $T(z)$ near $z=z_i$ equals
\be \label{texp}
T(z) =
\frac{1}{2\pi\kappa{\rm Tr}\left[L_1L_{-1}\right]} \left( \frac{{\rm Tr}\left[L_0 q_i\right]}{(z-z_i)^2} + 2 \frac{{\rm Tr}\bigl[q_i \hat{G}_0^{-1}\partial
\hat{G}_0(z_i)\bigr]}{z-z_i} + \ldots \right).
\ee
We therefore see that the residue $p_i$ at $z=z_i$ equals
\be \label{tres}
p_i = \frac{1}{\pi\kappa{\rm Tr}\left[L_1L_{-1}\right]} {\rm Tr}\bigl[q_i \hat{G}_0^{-1}\partial
\hat{G}_0(z_i)\bigr] = -\frac{1}{\pi\kappa{\rm Tr}\left[L_1L_{-1}\right]} \frac{\partial S_{\rm Toda}}{\partial z_i}
\ee
and for
\be
\kappa = \frac{c}{6\pi{\rm Tr}\left[L_1L_{-1}\right]}
\ee
this agrees precisely with (\ref{px}). In other words, derivatives of the semiclassical correlation functions of Toda theory do
give rise to the relevant first order pole in the expansion of the stress tensor. 

There are several other ways to obtain this result. One is to start with the the semiclassical approximation to the 
correlation function of a number of heavy operators and to expand the answer to first order in the conformal dimensions
of a subset of the operators. Another is to use the fact that one is computing the correlation function of a set of
primaries and use the Virasoro Ward identities.

\subsection{Semiclassical correlators in Toda revisited}
%%%%%%%%%%%%%%%%%%%%%%%%%%%%%%%%%%%%%%%%%%%%%%%%%%%%%
As we mentioned above, we have to be careful when doing an actual Toda computation, since the classical
saddle point for the computation involving the heavy operators may have a number of free parameters.
That such free parameters indeed exist is easy to see from (\ref{d0}): we can make redefinitions
$g(z)\rightarrow g(z) h$ and $\bar{g}(\bar{z}) \rightarrow
h' \bar{g}(\bar{z})$ with arbitrary $h$, $h'$ in (\ref{d0}).
This will generate an ambiguity in the solution of the Toda equations that
we can associate to the gauge fields $a$ and $\bar{a}\,$. In general this
ambiguity can be expressed as follows: for any $V\in SL(N,\mathds{C})$, and for a given saddle point $g(z)$, $\bar{g}(\bar{z})$
which solve (\ref{d0}), we can define a solution $G_0(V)$ of the Toda field equations through the decomposition
\be
\label{d1mod}
G(V)=g(z)V\bar{g}(\bar{z}) = G_-(V) G_0(V) G_+(V)\, .
\ee
The free parameters that we have in Toda theory are therefore given by an arbitrary $V\in SL(N,\mathds{C})$.
Accordingly, (\ref{todaresult}) is not quite true as stated, the right hand side should still involve an integral
over the $SL(N,\mathds{C})$ group element V
\be \label{todaresultmod}
e^{-S_{\rm Toda}} = \bigl\langle V_1(z_1)\ldots V_k(z_k)\bigr\rangle_{\rm semiclassical}
  \sim e^{-S_{\rm heavy}} \int {\cal D}V \,\prod_i{}' e^{{\rm Tr}\left[q_i \log \hat{G}_0(V;z_i)\right]}~,
\ee
where ${\cal D}V$ represents an $SL(N,\mathds{C})$-invariant measure.

%For $N=2$, we can parametrize the $SL(2,\mathds{C})$ matrix $V$ as
%\be
%V = \left( \begin{array}{cc} 1 & 0 \\ s_- & 1 \end{array} \right)
%\left( \begin{array}{cc} e^{s_0} & 0 \\ 0 & e^{-s_0} \end{array} \right)
%\left( \begin{array}{cc} 1 & s_+ \\ 0 & 1 \end{array} \right)
%\ee
%with $SL(2)$ invariant measure $\sim e^{2s_0}ds_0ds_+ds_-$. By explicit computation one obtains
%\be
%G_0 = \left(\begin{array}{cc} \psi & 0 \\ 0 & \psi^{-1} \end{array} \right),\qquad
%\psi^{-1} = \frac{e^{-s_0} z^{(1-\alpha)/2} }{\alpha} \left(-e^{2s_0} z^{\alpha} (\bar{z}+s_+)
%+\alpha (1+e^{2 s_0} s_-(s_+ +\bar{z})) \right) .
%\ee

We now specialize to the case of two light operators with charges $q_1$ and $q_2$ at locations $z_1$ and $z_2\,$.\footnote{To avoid cluttering, we will abuse notation and simply refer to the background solution $\hat G_0$ as $G_0\,$.}
The fields that enter in the integral $G_0(V;z_1)$ and $G_0(V;z_2)$ are given by
\be
g(z_i) V \bar{g}(\bar{z}_i) = G_-(V;z_i) G_0(V;z_i) G_+(V;z_i)~ .
\ee
Because the measure is invariant, we immediately see that the answer can only depend on the combinations
\be
X=g(z_1) g(z_2)^{-1}~, \qquad Y=\bar{g}_2(\bar{z}_2)^{-1} \bar{g}_1(\bar{z}_1)~.
\ee
Moreover, if we e.g. change $g(z_1)$ into $A_- g(z_1)$ with some constant $A_-\,$, then $A_-$
can be completely absorbed into $G_-(V;z_1)$ and will not affect $G_0(V;z_1)$ and hence also
not affect the integrand. With similar considerations for the other group valued fields, the integral
must be invariant under\footnote{Here ``$-$'' denotes upper triangular and ``$+$'' lower triangular, both with one's along the diagonal. }
\be \label{sym1}
X \rightarrow A_- X B_-~,\qquad  Y \rightarrow C_+ Y D_+ ~.
\ee
Finally, suppose that we multiply $g(z_i)$ by a constant diagonal matrix $A_0$ on the left.
By conjugating $A_0$ through $G_-(V;z_i)$, we see the only effect on the Toda field $G_0(V;z_i)$
is that it gets changed to $A_0 G_0(V;z_i)$. But then the integrand picks up a multiplicative
factor
\be
e^{-S_{\rm Toda}} \rightarrow e^{{\rm Tr}\left[q_i \log A_0\right]}  e^{-S_{\rm Toda}}\,  .
\ee
If we similarly consider multiplying $\bar{g}(\bar{z}_i)$ by a constant diagonal matrix from the
right, we find that in terms of $X$, $Y$ the following identies must hold
\bea \label{sym2}
X\rightarrow A_0 X & {\rm then} & e^{-S_{\rm Toda}} \rightarrow e^{{\rm Tr}\left[q_1 \log A_0\right]}  e^{-S_{\rm Toda}}~,
\nonumber \\
X\rightarrow X A_0 & {\rm then} & e^{-S_{\rm Toda}} \rightarrow e^{{\rm Tr}\left[-q_2 \log A_0\right]}  e^{-S_{\rm Toda}}~,
\nonumber \\
Y\rightarrow A_0 Y & {\rm then} & e^{-S_{\rm Toda}} \rightarrow e^{{\rm Tr}\left[-q_2 \log A_0\right]}  e^{-S_{\rm Toda}}~,
\nonumber  \\
Y\rightarrow Y A_0 & {\rm then} & e^{-S_{\rm Toda}} \rightarrow e^{{\rm Tr}\left[q_1 \log A_0\right]}  e^{-S_{\rm Toda}}~.
\eea
With these observations, we can completely determine the two-point function of light operators in a background
generated by heavy operators as we now illustrate for the case of $SL(2)$.

For $SL(2)$, it is easy to verify that $X_{21}$ and $Y_{12}$ are invariant under (\ref{sym1}), and that the two-point
function cannot depend on any of the other matrix entries of $X$ and $Y$. We denote the final answer
by $Z(X_{21},Y_{12})$, and the charges by $q_i={\rm diag}(q_i,-q_i)$.
The rescalings in (\ref{sym2}) then turn into
\be
Z(e^{\lambda}X_{21},Y_{12}) = e^{-2 q_1 \lambda}Z(X_{21},Y_{12}) = e^{-2 q_2 \lambda} Z(X_{21},Y_{12})~,
\ee
and
\be
Z(X_{21},e^{\lambda}Y_{12}) = e^{-2 q_2 \lambda}Z(X_{21},Y_{12}) = e^{-2 q_1 \lambda} Z(X_{21},Y_{12})~.
\ee
These equations only have a solution if $q_1=q_2\,$, which is indeed the case for which the light operators have the
same conformal dimension, and moreover we obtain
\be
Z = {\cal N}(X_{21}Y_{12})^{-2q_1}~,
\ee
where ${\cal N}$ is some normalization constant.
For the $SL(2)$ case, the group elements or fundamental matrices $g(z)$, $\bar{g}(\bar{z})$ can be chosen to
be equal to
\be
g(z) = \left( \begin{array}{cc} \frac{1+\alpha}{2\alpha} z^{\frac{-1+\alpha}{2}} &
-\frac{1-\alpha}{2} z^{\frac{-1-\alpha}{2}}
\\
-\frac{1}{\alpha}z^{\frac{1+\alpha}{2}} & z^{\frac{1-\alpha}{2}}  \end{array} \right) ~,
\quad
\bar{g}(\bar{z}) =  \left( \begin{array}{cc}  \frac{1+\alpha}{2\alpha} \bar{z}^{\frac{-1+\alpha}{2}}  & \frac{1}{\alpha}\bar{z}^{\frac{1+\alpha}{2}} \\ \frac{1-\alpha}{2}\bar{z}^{\frac{-1-\alpha}{2}} & \bar{z}^{\frac{1-\alpha}{2}}  \end{array} \right)~.
\ee
Setting $z_1=x$ and $z_2=1$ with  $x$ real, we find
\be
X_{21} =  \frac{x^{\frac{1-\alpha}{2}} - x^{\frac{1+\alpha}{2}}}{\alpha}~,\quad Y_{12}=-\frac{x^{\frac{1-\alpha}{2}} - x^{\frac{1+\alpha}{2}}}{\alpha} ~.
\ee
We finally get
\be
Z = {\cal N} \left( \frac{\alpha^2}{ (x^{\frac{1-\alpha}{2}} - x^{\frac{1+\alpha}{2}})^2 } \right)^{2q_1} ~.
\ee
 which agrees perfectly with (\ref{correlator1}) obtained using the monodromy method.

It is interesting to see that the Toda computation involves the matrices $X$ and $Y$, which are also the main
building blocks of the Wilson loop computation. It would be interesting to prove directly that the Toda computation
and the Wilson loop computations agree.

It turns out that symmetries are also sufficient to compute the two-point function in the $SL(N)$ case. One can prove
that the following variables
\be
X^{[p]}= {\rm det}_{N-p+1 \leq i \leq N,1\leq j\leq p}(X_{ij})~, \qquad
Y^{[p]}= {\rm det}_{1\leq i \leq p,N-p+1 \leq j \leq N} (Y_{ij})~,
\ee
are the only quantities we can make out of the $SL(N)$ matrices $X$ and $Y$ which are invariant under (\ref{sym1}).
The Toda correlation function can therefore only be a function of these variables.

Repeating the same arguments as in the $SL(2)$ case, we can determine the semiclassical correlation function for
arbitrary $N$. If we denote $ q_i = {\rm diag}(q_i)_k\,$, then the two point function of light operators is only
non-vanishing if
\be
(q_2)_i = -( q_1)_{N+1-i}~,
\ee
and if this condition is satisfied the correlation function equals
\be \label{mainres}
Z={\cal N} \prod_{p=1}^{N-1} (X^{[p]}Y^{[p]})^{( q_1)_{N+1-i}-( q_1)_{N-i}}~.
\ee

Equation (\ref{mainres}) is the main result of our Toda computation, and it expresses
the correlation function explicitly in data determined by the background. In principle,
the same methods could be used to analyze higher point functions, and it would be interesting
to explore this in more detail. As we mentioned above, it would also be worthwhile to compare
this result to both the Wilson line computation as well as to the monodromy computation.
Our derivation has perhaps been somewhat heuristic, as it relied on scaling arguments
based on an integral over the non-compact group $SL(N,\mathds{C})$. This group has infinite
volume and a more careful treatment of this integral would be desirable. It is also not entirely
clear to us whether one should actually do the full integral or choose a suitable real slice, which
is related to the fact that most of our discussion relied on a complexification of Toda theory
whose precise interpretation also requires further clarification.

Something which deserves a further explanation is why our computation appears to pick out the
vacuum block. We suspect that the integral over $V$ plays a crucial role here. It is tempting 
to speculate that the integral over $V$ projects the intermediate channel between the heavy and
the light operators onto the identity operator, and that one might be able to obtain the contributions
of other blocks by inserting a suitable $SL(N,\mathds{C})$ character into the path integral. We hope
to return to this issue in the future.  

As an aside, we notice that it is relatively straightforward to analyze single-valuedness of a Toda solution from
the point of view of monodromies. Consider a solution of the Toda field equations given by $g(z)V\bar{g}(\bar{z})=
G_-G_0G_+\,$. If the Toda field $G_0$ is regular when going around a point $z_i$ where $g(z)\rightarrow g(z)M_i$
and $\bar{g}(\bar{z})\rightarrow \bar{M}_i \bar{g}(\bar{z})$ then $G_-G_-G_+$ must be single valued as well, since
$G_-$ and $G_+$ are local in terms of $G_0\,$. Therefore a necessary condition for single-valuedness is that for all $i$
\be
g(z) V\bar{g}(\bar{z}) = g(z)M_i V \bar{M}_i \bar{g}(\bar{z})~,
\ee
which is equivalent to
\be
V=M_i V \bar{M}_i~,
\ee
for all $i\,$. In particular, the background solution generated by a  heavy chiral operators, e.g. $\bar{M}_0=1$ but
$M_0\neq 1\,$,  does not correspond to a single valued Toda field. We have ignored this fact in our computation
and further work is required to determine the implications of this observation for the complexified theory.

%%%%%%%%%%%%%%%%%%%%%%%%%%%%%%%%%%%%%%%%%%%%%%%%%%%%%%%%%%%%%%%%%%%%%%%%%%%%%%%%
\section{Discussion}\label{discussion}
%%%%%%%%%%%%%%%%%%%%%%%%%%%%%%%%%%%%%%%%%%%%%%%%%%%%%%%%%%%%%%%%%%%%%%%%%%%%%%%%
We end our work by discussing some important features of our results and some possible future directions.

%%%%%%%%%%%%%%%%%%%%%%%%%%%%%%%%%%%%%%%%%%%%%%%%%%%%%
\subsection{Microstates versus effective geometries}
%%%%%%%%%%%%%%%%%%%%%%%%%%%%%%%%%%%%%%%%%%%%%%%%%%%%%

It is interesting to think more about the meaning of the agreement between the bulk and CFT computations presented here. Recall that on the CFT side we are computing a contribution to a vacuum four-point function, or equivalently a two-point function evaluated in an excited state. The excited state is one that is produced by acting with a heavy local operator on the vacuum. We are only keeping the leading large $c$ part of the vacuum (Virasoro or ${\cal W}_3$) block contribution to these correlation functions. As we have found, this CFT result is reproduced by computing the action of a probe particle moving in a background solution whose charges correspond to those of the heavy CFT operator. The large $c$ approximation corresponds to treating the probe and background classically, and the restriction to the vacuum block corresponds to including only massless higher spin fields in the bulk, and not additional matter fields.

As discussed in \cite{Fitzpatrick:2014vua}, the above story is particularly interesting when the charges carried by the background are such that we are in the black hole regime. If we turn off the spin-3 charges so that we have a pure metric solution in the bulk, we recall that a BTZ black hole is obtained by taking the conformal dimension of the heavy operator to obey $h, \overline{h} > {1\over 4}\,$. The BTZ black hole solution is usually thought of as describing a system in thermal equilibrium; for example, the correlation functions computed in this background will be periodic in imaginary time, indicating a well defined temperature. On the other hand, the CFT computation that we are comparing to makes reference to a specific microstate, not a thermal ensemble. Apparently, upon taking the large c limit and restricting to the vacuum block, the microstate has been replaced by an effective thermal ensemble. This type of phenomenon has been discussed before in the AdS/CFT correspondence (see \cite{Balasubramanian:2005mg,Balasubramanian:2005qu}) and clearly has bearing on the black hole information paradox.

With the results found here, we can ask how the story changes when we include higher spins. In particular, we can ask whether for sufficiently large conformal dimension the effective background solution is a higher spin black hole. We first address when we would expect a black hole interpretation to be appropriate. Recall that on the cylinder the correlation function is built out of combinations of $e^{ip_n w}$, where $p_{1,2,3}$ are the three distinct roots of the cubic $p_n^3 -(1-4h_2)p_n-4q_2=0\,$. For real $p_n$ these exponentials are oscillatory for real time on the cylinder, while they grow/decay exponentially if $p_n$ acquire an imaginary part. Imaginary parts occur for $h_2 > h_2^{\rm crit}$, where $h_2^{\rm crit} = {1\over 4} - ({108 \over 64}q_2^2)^{1/3}\,$. For $h_2 > h_2^{\rm crit}$, it is then easy to see that the correlation function on the cylinder will decay to zero at large real time. This behavior is what one expects in the presence of an event horizon, with the infinite redshift at the horizon being responsible for the exponential decay. A solution with a mass gap would instead lead to oscillatory behavior.

This conclusion can also be reached by examining the holonomy of the connection around the angular direction. As first shown in \cite{deBoer:2013gz}, the entropy of a higher spin black hole can be written $S=2\pi k_{\rm cs} {\rm Tr}[L_0(\lambda_\phi-\overline{\lambda}_\phi)]$, where $\lambda_\phi$ is a diagonal matrix whose entries are the eigenvalues of $a_\phi\,$. Clearly this identification requires the eigenvalues to be real, and a quick computation shows that this requires $h_2 > h_2^{\rm crit}$, as above.

However, two closely related facts make the identification with a higher spin black hole more subtle than in the BTZ case. First, a proper higher spin black hole solution should have trivial holonomy around a Euclidean time circle. For this to be the case, the connection needs to have both $a_w$ and $a_{\wb}$ turned on, whereas we have seen that the CFT result matches on to a connection with only $a_w\,$. %Turning on $a_{\wb}$ would apparently spoil the agreement with the CFT. 
Turning on $a_{\wb}$ would require introducing sources (chemical potentials) in the CFT computation, thus deforming the CFT Hamiltonian. Second, the correlators we have computed are not periodic in imaginary time, as would be expected for a thermal interpretation. This can be seen from the fact that the $p_n$ are not rational multiples of each other, and is also a consequence of the lack of trivial holonomy around a thermal circle.

To interpret this, consider the simpler situation of a charged black hole in Einstein-Maxwell theory. Usually, one sets $A_t=0$ at the horizon, so that $A_\mu$ is a well defined vector field on the Euclidean geometry. Doing so, correlation functions of fields exhibit thermal periodicity. Suppose one instead applies a constant shift to $A_t$ so as to set $A_t=0$ at infinity. In this case, correlation functions of charged fields will not exhibit thermal periodicity, as is easily seen by noting that the gauge transformation that relates the two cases is not single valued around the thermal circle. Our higher spin background with $a_{\wb}=0$ is analogous to the Einstein-Maxwell black hole with $A_t=0$ at infinity. To obtain the usual higher spin black hole we should perform a non-single valued higher spin gauge transformation. This bulk gauge transformation should be accompanied by a corresponding finite ${\cal W}_3$ transformation acting on the operators in the CFT so as to maintain agreement between the bulk and boundary correlators. Carrying out this transformation explicitly is rather cumbersome, but the point is that, suitable interpreted, our computations are consistent with emergence of a higher spin black hole solution.

%%%%%%%%%%%%%%%%%%%%%%%%%%%%%%%%%%%%%%%%%%%%%%%%%%%%%
\subsection{Multiple intervals and higher genus boundary geometries}
%%%%%%%%%%%%%%%%%%%%%%%%%%%%%%%%%%%%%%%%%%%%%%%%%%%%%

In this paper we considered the entanglement entropy for a single interval on the plane; let us briefly comment on the more general case.  As we discussed, for $N_I$ intervals one should introduce $2N_I$ twist operators, so that the excited state entanglement entropy is captured by a $2N_I+2$-point correlation function.    The monodromy analysis has to be extended accordingly.  In the expressions for $T(z)$ and $W(z)$ we should allow for poles at the locations of all the twist operators,
\begin{align}\label{modified currents}
T(z)
={}&
\sum_{i = 1}^{2N_{I}}\frac{h_{i}}{(z-z_{i})^{2}} + \frac{p_{i}}{z-z_{i}}\,,
\\
W(z)
={}&
\sum_{i = 1}^{2N_{I}}\frac{q_{i}}{(z-z_{i})^{3}}+\frac{a_{i}}{(z-z_{i})^{2}} + \frac{b_{i}}{z-z_{i}}  \,.
\label{modified currents 2}
\end{align}
Note that in writing this we have made the important assumption of replica symmetry, which implies that $T(z)$ and $W(z)$ should be single valued in $z$.   If we relax the condition of replica symmetry, then nothing would stop us from adding additional holomorphic quadratic and cubic differentials\footnote{More properly meromorphic differentials in the presence of additional insertions such as the operators creating an excited state.} to the right hand side.  Assuming replica symmetry, we are still left with the challenging problem of fixing the accessory parameters by imposing trivial monodromy around various cycles.  However, if we are only interested in entanglement entropy rather than R\'enyi entropy, the problem is a rather trivial extension of the single interval case.  Recall that for entanglement entropy we work to first order in $\varepsilon \sim n-1\,$.     At first order there is no crosstalk between distinct intervals, and so the solution is found from superposition.   This point was emphasized in \cite{Faulkner:2013yia,Hartman:2013mia}, and of course agrees with the Ryu-Takayanagi formula in the Virasoro case.  In our case, we will get agreement with the Wilson line results if we simply take multiple Wilson lines connecting the various endpoints in pairs.  The correct pairing of endpoints depends on the locations of the twist operators, and there can be phase transitions as these are varied; again, see \cite{Faulkner:2013yia,Hartman:2013mia} for more discussion.   For the R\'enyi entropy, there will be a more intricate interplay between the distinct intervals.

Replacing the plane by a higher genus Riemann surface also introduces new aspects that could be interesting to consider. The case of the torus is of particular relevance due to its thermal interpretation, and is related to the discussion of black holes in the previous subsection.   In the monodromy analysis, the large $z$ falloff conditions on $T(z)$ and $W(z)$ will be replaced by periodicity conditions around the nontrivial cycles. For a single interval on the torus, this  problem was addressed in the Virasoro case in \cite{Barrella:2013wja}. Unless $W(z)=0$, in $SL(N)$ Chern-Simons theory the periodicity along the thermal cycle requires one to reintroduce  the $a_{\bar z}$ and $\bar a_z$ components of the connections. The currents $T$ and $W$ will no longer be holomorphic, and our ODE might turn into an unpleasant PDE. Still  the problem of constructing a regular connection supported by this background should be doable. In the CFT it is not evident that we must modify drastically our currents. It would be interesting to realize  the bulk conditions of the CS connections as constraints for the $n$-point functions on the torus for a  ${\cal W}_N$ CFTs .

On a general Riemann surface, the general ansatz for $T(z)$ will include a sum over holomorphic quadratic differentials with free coefficients, and likewise for $W(z)$.    It would be interesting to verify that the appropriate monodromy conditions uniquely fix all coefficients in the problem.

%%%%%%%%%%%%%%%%%%%%%%%%%%%%%%%%%%%%%%%%%%%%%%%%%%%%%%%%%%%%%%%%%%%%%%%%%%%%%%%%
\section*{Acknowledgements}
%%%%%%%%%%%%%%%%%%%%%%%%%%%%%%%%%%%%%%%%%%%%%%%%%%%%%%%%%%%%%%%%%%%%%%%%%%%%%%%%
It is a pleasure to thank Marco Baggio, Matthias Gaberdiel, Manuela Kulaxizi, Wei Li, Eric Perlmutter and Matteo Rosso  for discussions.  J.I.J. would also like to thank the participants of the conference ``Recent developments in String Theory" in Ascona for stimulating discussions, and the University of Krakow, AEI Potsdam and King's College for hospitality while this work was in progress. A.C. is supported by Nederlandse Organisatie voor Wetenschappelijk Onderzoek (NWO) via a Vidi grant.  The work of J.I.J. is partially supported by the Swiss National Science Foundation and the NCCR SwissMAP. P.K. is supported in part by NSF grant PHY-1313986.  This work was  as well  supported in part by the National Science Foundation under Grant No. PHYS-1066293 and the hospitality of the Aspen Center for Physics.

\appendix

%%%%%%%%%%%%%%%%%%%%%%%%%%%%%%%%%%%%%%%%%%%%%%%%%%%%%%%%%%%%%%%%%%%%%%%%%%%%%%%%
\section{Conventions}
%%%%%%%%%%%%%%%%%%%%%%%%%%%%%%%%%%%%%%%%%%%%%%%%%%%%%%%%%%%%%%%%%%%%%%%%%%%%%%%%
Here we collect some useful formulas for handy reference. The Chern-Simons action is
%
%\be\label{c1}
%I_{\rm CS} = \frac{k_{\rm cs}}{4\pi} \int_M\! {\rm Tr} \left[AdA + \frac{2}{3}A^3\right] - \frac{k_{\rm cs}}{4\pi} \int_M\! {\rm Tr} \left[\Ab d\Ab + \frac{2}{3}\Ab^3 \right]
%\ee
\begin{align}\label{sl(N) CS action}
I_{\rm CS} \equiv{}&
 \frac{k_{\rm cs}}{4\pi}\int_{M} \mbox{Tr}\Bigl[CS(A) - CS(\overline{A})\Bigr]
\end{align}

\noindent where
\begin{equation}
CS(A) \equiv AdA + \frac{2}{3}A^3\,.
\end{equation}

\noindent The bulk Newton constant is related to the central charge and the Chern-Simons level as
\be
c = {3\ell\over 2 G}=12 \Tr \left[L_0 L_0\right] k_{\rm cs}~.
\ee
The generalized vielbein and metric are
\be e = {1\over 2}(A-\Ab)~,\quad  g_{\mu\nu} =  {1\over \Tr \left[L_0 L_0\right]} \Tr \left[e_\mu e_\nu\right]~.
\ee
\noindent The $sl(N)$  generators are defined as in \cite{Castro:2011iw}.  In particular, for $sl(3)$ we have\footnote{In a slight abuse of notation, in the main text we have also used the symbols $L_{n}$, $W_{m}$ to denote the modes of the $\mathcal{W}_{3}$ algebra \eqref{W3 algebra commutators}. We trust that the intended meaning should be clear from the context.}
%%
%\bea
%L_{1}&=& -\sqrt{2}\left( \begin{array}{ccc}0&0&0\\1&0&0\\0&1&0\\ \end{array}\right)~, \quad
%L_{0}= \left( \begin{array}{ccc}1&0&0\\0&0&0\\0&0&-1\\ \end{array}\right)~, \quad
%L_{-1}= \sqrt{2}\left( \begin{array}{ccc}0&1&0\\0&0&1\\0&0&0\\ \end{array}\right)~, \cr && \cr
%W_{1}&=& -{1\over{\sqrt{2}}}\left(  \begin{array}{ccc}0&0&0\\
% 1&0&0\\
%0&-1&0\end{array}\right)~,\quad
%W_{0}={1\over 3}\left(  \begin{array}{ccc}1&0&0\\
% 0&-2&0\\
%0&0&1\end{array}\right)~,\quad
%W_{-1}={1\over \sqrt{2}}\left(  \begin{array}{ccc}0&1&0\\
% 0&0&-1\\
%0&0&0\end{array}\right)~,\cr && \cr
%W_{2}&=&2\left(  \begin{array}{ccc}0&0&0\\
% 0&0&0\\
%1&0&0\end{array}\right)~, \quad
%W_{-2}=2\left(  \begin{array}{ccc}0&0&1\\
% 0&0&0\\
%0&0&0\end{array}\right)
%\eea
%%
\begin{align}
L_{1} &=
-\sqrt{2}\left(\begin{array}{ccc}
0 & 0 & 0 \\
1 & 0 & 0 \\
0& 1 & 0
\end{array}
\right),&
L_{0} &=
\left(\begin{array}{ccc}
1 & 0 & 0 \\
0 & 0 & 0 \\
0& 0 & -1
\end{array}
\right),&
L_{-1} &=
\sqrt{2}\left(\begin{array}{ccc}
0 & 1 & 0 \\
0& 0 & 1\\
0 & 0 & 0
\end{array}
\right),
\nonumber\\
W_{2} &=
 2
\left(
\begin{array}{ccc}
0& 0 & 0 \\
0 & 0 & 0\\
1& 0 & 0
\end{array}
\right),&
W_{1} &=
-\frac{1}{\sqrt{2}}\left(
\begin{array}{ccc}
0& 0 & 0 \\
1 & 0 & 0\\
0& -1 & 0
\end{array}
\right),&
W_{0} &= \frac{1}{3}
\left(
\begin{array}{ccc}
1& 0 & 0 \\
0 & -2 & 0\\
0& 0 & 1
\end{array}
\right)
,&
\\
W_{-1} &=
\frac{1}{\sqrt{2}}
\left(
\begin{array}{ccc}
0& 1 & 0 \\
0 & 0 & -1\\
0& 0 & 0
\end{array}
\right),&
W_{-2} &=
2
\left(
\begin{array}{ccc}
0& 0 & 1 \\
0 & 0 & 0\\
0& 0 & 0
\end{array}
\right).
\nonumber
\end{align}

The commutation relations then  read
\bea
[L_m,L_n]&=&(m-n)L_{m+n}~,\cr
[L_m,W_n]&=&(2m-n)W_{m+n}~,\cr
[W_m,W_n]&=&-{1\over{12}}(m-n)(2m^2+2n^2-mn-8)L_{m+n}~.
\eea
The connection corresponding to the Euclidean BTZ solution is (where, as is standard, we have gauged away the dependence on the radial coordinate)
\bea
a&=& (L_1 - P L_{-1})dw ~,\cr
\ab&=& (L_{-1} -\overline{P} L_1)d\wb~,
\eea
and the metric is
\be
ds^2 = d\rho^2 + P dw^2 + \overline{P} d\wb^2 +\left(e^{2\rho} +P \overline{P} e^{-2\rho}\right)dwd\wb~.
\ee
Here $w = \phi+it$, and $\wb = \phi-it\,$.   The components of the conventionally normalized CFT stress tensor are
\be
T_{CFT}(w) = -{c\over 6} P ~,\quad \Tb_{CFT}(\wb) = -{c\over 6}\overline{P}~.
\ee
The Virasoro zero modes are
\be
L_0 ={c\over 6} P +{c\over 24}~,\quad \Lb_0 = {c\over 6}\overline{P}_{CFT} +{c\over 24}~.
\ee
The BTZ solutions have $P, \overline{P} \geq 0\,$.   Conical defects have $-{1\over 4} < P, \overline{P} < 0\,$.
The stress tensor on the z-plane, $z=e^{iw}$, is given by
\be
T_{CFT}(w) = -z^2 T_{CFT}(z) +{c\over 24}~,\quad \Tb_{CFT}(\wb) = -\zb^2 \Tb_{CFT}(\zb)+{c\over 24}~.
\ee
It will be convenient to pull out a factor of $c/6$ from the definition of the stress tensor and define
\be
T_{CFT} = {c\over 6}T~,\quad  \Tb_{CFT} = {c\over 6}\Tb~.
\ee
With this in mind, for SL(3) we will write the connections on the plane as
\bea
a & =& \Bigl(L_1 + T(z)L_{-1} + W(z)W_{-2}\Bigr)dz~, \cr
\ab & =& \Bigl(L_{-1} + \Tb(\zb)L_{1} + \Wb(\zb)W_{2}\Bigr)d\zb~.
\eea
An operator at the origin with charges $(h,q)$ will correspond to
\be
T(z) = {h\over z^2}~,\quad W(z) = {q\over z^3}~,\quad  \Tb(\zb) = {\hb\over \zb^2}~,\quad \Wb(\zb) = {\qb \over z^3}~.
\ee
Transforming to the cylinder via $z=e^{iw}$ then gives
\be
T(w) =-h+{1\over 4}~,\quad W(w) = -iq~,\quad \Tb(\wb) = -\hb+{1\over 4}~,\quad \Wb(\wb) =i\qb~.
\ee
At $q=\qb=0\,$, conical defects have $0 < h,\hb < {1\over 4}\,$; and BTZ solutions have $ h,\hb > {1\over 4}\,$. On the cylinder the connections of course have the same basic form as on the plane,
\bea
a & =& \Bigl(L_1 + T(w)L_{-1} + W(w)W_{-2}\Bigr)dw ~,\cr
\ab & =& \Bigl(L_{-1} + \Tb(\wb)L_{1} + \Wb(\wb)W_{2}\Bigr)d\wb~.
\eea
%

%%%%%%%%%%%%%%%%%%%%%%%%%%%%%%%%%%%%%%%%%%%%%%%

%%%%%%%%%%%%%%%%%%%%%%%%%%%%%%%%%%%%%%%%%%%%%%%

\section{Spin-3 Entropy}\label{app:s3}

In this appendix we will discuss some aspects of spin-3 entanglement and thermal entropy as defined in \cite{Hijano:2014sqa}. The microscopic definition of these entropies is still rather unclear. Our aim here is to  investigate some properties of the bulk definitions which could give further guidance to a proper boundary CFT definition.

%%%%%%%%%%%%%%%%%%%%%%%%%%%%%%%%%%%%%%%%%%%%%%%%%%%%%
\subsection{Generalized R\'enyi entropies}
%%%%%%%%%%%%%%%%%%%%%%%%%%%%%%%%%%%%%%%%%%%%%%%%%%%%%
For the purpose of computing R\'enyi entropies, the quantum numbers of the (anti-)twist operator is fixed by demanding that it captures the correct geometric data of the problem. In the canonical definition of R\'enyi entropy, given by \eqref{Renyi from partition function} and \eqref{excited Z}, the twist operators encode the data of the branch cuts in the replicated geometry. As explained in  section \ref{subsubsec: branch cuts}, the equation that determines the conformal dimension of the twist operator is
\begin{equation}\label{eq:mi3}
{\rm eigenvalues}\Bigl[\left(M_{i}\right)^{n}\Bigr] =\pm {\rm eigenvalues}\Bigl[{e^{2\pi i L_0}}\Bigr]\,,
\end{equation}
which gives  that the (anti-)twist operator has weight
\be
\Delta = \frac{nc}{6}h=\frac{c}{24}\left(n-\frac{1}{n}\right)~.
\ee

In the presence of extended algebras, such as ${\cal W}_N$, it is rather natural to design a ``new twist operator''  that carries quantum numbers associated to the additional higher spin conserved currents \cite{Hijano:2014sqa}. And along the lines  of the derivations in \eqref{excited Z}, it is tempting to give  a geometrical interpretation to this new twist. For concreteness, we focus on $N=3$. In this case, we know that by imposing regularity of a bulk Wilson line \cite{Hijano:2014sqa} the quantum numbers of the spin-3 twist are at leading order
\be\label{eq:hq3}
h=O(n-1)^2 ~,\quad {nc\over 6}q=-{c\over 12}(n-1)+O(n-1)^2~.
\ee
Since the operator is charged under the spin 3 current, it seems like we are inducing a ``branch cut'' via a spin-3 gauge transformation (whatever this means!). It  seems reasonable  to then generalize the r.h.s. of \eqref{eq:mi3} so that  we can accommodate the charges in \eqref{eq:hq3}. Writing  \eqref{eq:mi3} as
\be
\exp \bigl(2\pi i L_0\bigr) = \exp \le(2\pi i n \left(L_0+ {1-n\over n}L_0\right)\ri)~,
\ee
a reasonable generalization is to impose
\be\label{eq:wl}
{\rm eigenvalues}\Bigl[\left(M_{i}\right)^{n}\Bigr] = {\rm eigenvalues} \le[\exp\le( 2\pi i n \left(L_0+{3}{1-n\over n}W_0\right)\ri)\ri]
\ee
This combination of matrices has the feature that in the limit $n\to 1$ we would reproduce \eqref{eq:hq3}, and the r.h.s. is in the center of $SL(3)$. It will as well nicely fit with the thermal  $S_3$ entropy (discussed below). However, beyond being a simple and elegant choice, \eqref{eq:wl} is not unique. The leading terms in \eqref{eq:hq3} do not provide enough data to unambiguously determine the condition on the monodromy matrix at finite $n\,$. To either confirm or refute \eqref{eq:wl} we need to understand what is the  geometrical interpretation\footnote{In this context, a geometrical interpretation will likely require treating spin-3 gauge transformation and diffeomorphisms in an equal footing.} of the spin-3 twist fields.

%%%%%%%%%%%%%%%%%%%%%%%%%%%%%%%%%%%%%%%%%%%%%%%%%%%%%
\subsection{Thermal $S_3$}
%%%%%%%%%%%%%%%%%%%%%%%%%%%%%%%%%%%%%%%%%%%%%%%%%%%%%

In this subsection we will show how to obtain the generalized thermal entropy of \cite{Hijano:2014sqa} from an Euclidean Chern-Simons action.

The Euclidean Chern-Simons action for a general pair of Drinfeld-Sokolov connections carrying zero modes (namely charges and their conjugate chemical potentials) on the torus with identifications $z\cong z+2\pi \cong z+2\pi \tau$ was computed  in \cite{deBoer:2013gz,deBoer:2014fra}. By performing a Legendre transformation, the thermal entropy of the system is found to be
\begin{align}\label{eq:gs}
S &= -2\pi i k_{cs}\text{Tr}\Bigl[\left(a_{z} + a_{\bar{z}}\right)\left(\tau a_{z} + \bar{\tau}a_{\bar{z}}\right)\Bigr] + \text{barred}
\\
&=
-2\pi i k_{cs}\text{Tr}\left[a_{\phi}h\right] + \text{barred}\,,
\end{align}
where
\begin{equation}
a_{\phi}\equiv a_{z} +a_{\bar{z}}\,,\qquad h \equiv \tau a_{z} +\bar{\tau}a_{\bar{z}}\,.
\end{equation}
We emphasize that  the form of the connection, variational principle and boundary terms  remain exactly the same as for the derivations in \cite{deBoer:2013gz,deBoer:2014fra}. The main difference comes about in regularity condition of the connections around the thermal cycle. We propose that
\begin{equation}\label{spin 3 smoothness conditions}
\text{spin-3 smoothness:}\qquad  \text{eigenvalues}\left[h\right]= \text{eigenvalues}\left[3iW_0\right],
\end{equation}
as opposed to $ \text{eigenvalues}\left[h\right] =   \text{eigenvalues}\left[iL_0\right]$ which is the smoothness condition that yields the usual thermal entropy. This new smoothness condition is compatible with the conditions imposed on the Wilson line \cite{Hijano:2014sqa}. It as well seems compatible with the condition imposed on the branch cuts for the generalized spin-3 R\'enyi entropy \eqref{eq:wl}.

We will use canonical boundary conditions, which map to deformations of the Hamiltonian in the dual CFT. As explained in detail in \cite{deBoer:2014fra}, this means the charges sit in $(a_{z}+a_{\bar{z}})$ and their conjugate potentials in $a_{\bar{z}}\,$. More precisely, we consider the following constant flat $sl(3)$ connection:
\begin{align}
a_{\phi} &= a_{z} +a_{\bar{z}}
= L_1 - {6\over c}\mathcal{L}L_{-1} - {6\over c}\mathcal{W}W_{-2}~,
 \\
a_{\bar{z}}
&=-
 \frac{ \nu_{3}}{2}\left(a_{\phi}^{2} - \frac{1}{3}\text{Tr}\left[a_{\phi}^{2}\right]\mathds{1}\right)~.
\end{align}
For simplicity, let us consider the non-rotating case, and define an inverse ``spin-3 temperature" $\beta_{3}$ through
\begin{equation}
\tau = - \bar{\tau}  = \frac{i\beta_{3}}{2\pi}\,.
\end{equation}
The smoothness conditions \eqref{spin 3 smoothness conditions} reduce to
\begin{equation}\label{eq:h3}
\det \left[a_{z} -a_{\bar{z}}\right] =- \frac{16\pi^{3}}{\left(\beta_{3}\right)^{3}}~,\qquad \text{Tr}\left[\left(a_{z} -a_{\bar{z}}\right)^{2}\right] = \frac{24\pi^{2}}{\left(\beta_{3}\right)^{2}}\,.
\end{equation}
In contrast to the usual definition of smoothness, we can solve for the charges in terms of the potentials in a simple manner. The solution to \eqref{eq:h3} is
\bea\label{sol spin 3 smoothness 1}
{6\over c}\mathcal{L} &=&
 {3\over \left(2\nu_{3}\right)^{2}}\left(1+2\pi \frac{\nu_{3}}{\beta_{3}}\right)~,
 \cr
{6\over c}\mathcal{W} &=&
\frac{1}{2\left(\nu_{3}\right)^{3}} \left(1+3\pi \frac{\nu_{3}}{\beta_{3}}\right)~.
\eea

Using the spin-3 smoothness condition \eqref{spin 3 smoothness conditions} in \eqref{eq:gs} we find that the spin-3 thermal entropy is
\begin{align}\label{spin 3 entropy via eigenvalues}
S_3 &= -2\pi i k_{\rm cs}\text{Tr}\Bigl[\left(a_{z} + a_{\bar{z}}\right)\left(\tau a_{z} + \bar{\tau}a_{\bar{z}}\right)\Bigr] + \text{barred}
\\
&=
6\pi k_{\rm cs}\text{Tr}\left[W_{0}\lambda_{\phi}\right] + \text{barred}\,,
\label{spin 3 entropy via eigenvalues}
\end{align}
This is an expression for the entropy as a function of the charges $({\cal L},{\cal W})$ and it agrees with the results in \cite{Hijano:2014sqa}. However, the smoothness condition \eqref{spin 3 smoothness conditions} gives a different relation between charges and potentials. In particular the first law in terms of these new definitions is
\begin{equation}\label{spin 3 first law}
\delta S_{3} = 2\pi i \left(\tau_{3}\delta \mathcal{L} + \alpha_{3}\delta \mathcal{W}\right) + {\rm barred}~,
\end{equation}
where we have defined
\begin{equation}\label{thermal potentials}
\tau_{3} = \frac{i\beta_{3}}{2\pi}~,\qquad \alpha_{3} = \frac{i}{\pi}\beta_{3}\nu_{3}~,
\end{equation}
This shows  consistency (integrability) for our new definition of potentials. Summarizing, a linear ``spin-3 first law" \eqref{spin 3 first law} is satisfied with thermal potentials given by \eqref{thermal potentials}, which as we have seen follow from the smoothness conditions \eqref{sol spin 3 smoothness 1}.

We can also define the ``spin-3 free energy" or spin-3 grand-canonical potential, which is the Legendre transform of the spin-3 entropy. Quoting the formula from \cite{deBoer:2014fra}
\begin{align}
\ln Z_{3}
&= -2\pi i k_{\rm cs} \text{Tr}\left[\frac{\tau}{2}\left(a_{z} + a_{\bar{z}}\right)^{2} + \left(\bar{\tau}-\tau\right)L_1 a_{\bar{z}} + \text{barred}\right]~,
\end{align}
and using the above solution of the smoothness conditions we find a very simple expression:
\begin{equation}
\ln Z_{3} =
-\frac{k_{\rm cs}}{\nu_{3}}\left(6\pi + \frac{\beta_{3}}{\nu_{3}}\right) + \text{barred}~.
\end{equation}

%%%%%%%%%%%%%%%%%%%%%%%%%%%%%%%%%%%%%%%%%%%%%%%

%%%%%%%%%%%%%%%%%%%%%%%%%%%%%%%%%%%%%%%%%%%%%%%

\section{Resonant monodromy}\label{subsubsec: resonant}
%\textcolor{red}{AC: add why this is useful.}
Branch cuts introduce resonant singular points in the ODE. They can also occur if the  charges of the heavy operators are tuned appropriately. In this case $\Phi$ is not invertible and some of our steps should be revisited. In this appendix we will elaborate more on the properties of the monodromy matrix for this peculiar case, which could be useful for future work.

In order to exemplify the significance of logarithmic branches of solutions and global properties of the monodromy around singular points, consider the $sl(2)$ case \eqref{eq:ode2} with
\begin{equation}\label{stress tensor ansatz}
T(z) = \frac{h}{z^{2}} + \frac{p}{z}\,,
\end{equation}

\noindent where, without loss of generality, we have chosen the singularity to be at $z=0\,$. The relevant ODE is then
\begin{equation}\label{our ODE}
\psi''(z) + \left(\frac{h}{z^{2}} + \frac{p}{z}\right)\psi(z) =0\,.
\end{equation}

\noindent It will prove convenient to define the quantity $\nu$ through
 \begin{equation}
 h = \frac{1}{4}\left(1-\nu^{2}\right),
 \end{equation}

\noindent in terms of which the roots of the indicial equation around $z=0$ are
\begin{equation}
 \Delta_{\pm} = \frac{1}{2} \pm \frac{\nu}{2} \qquad \Rightarrow \qquad \Delta_{+} - \Delta_{-} = \nu\,,
 \end{equation}

\noindent which implies that the cases where $\nu$ is an integer will generically admit logarithmic branches in the solution.

Let us first briefly revisit the non-resonant case, i.e. $ \nu \notin \mathds{Z}\,$.  In this case, the general solution of \eqref{our ODE} can be taken to be
\begin{equation}
 \psi(z) = c_{+}\,\sqrt{pz}\, J_{\nu}\bigl(2\sqrt{pz}\bigr) + c_{-}\,\sqrt{pz}\,J_{-\nu}\bigl(2\sqrt{pz}\bigr).
 \end{equation}
\noindent Using standard properties of Bessel functions we can follow the solution as $z \to z e^{2\pi i}$, and from \eqref{eqn:monodef} we read off
\begin{eqnarray}\label{non resonant monodromy}
M_{\gamma} =
 -\left(
\begin{array}{cc}
e^{i\pi \nu} & 0 \\
0 & e^{-i\pi \nu}
\end{array}
\right) =
 -\left(
\begin{array}{cc}
e^{i\pi \sqrt{1-4h}} & 0 \\
0 & e^{-i\pi \sqrt{1-4h}}
\end{array}
\right)
\end{eqnarray}

\noindent in agreement with \eqref{eq:mi2}. Not surprisingly, in the generic case the monodromy matrix is diagonalizable and it only depends on the leading singular behavior close to $z=0$ (namely on $h$).

 Suppose we are now in the resonant case
\begin{equation}
\nu = m \in \mathds{Z}
\end{equation}

\noindent instead. A basis of linearly-independent solutions of \eqref{our ODE} in this case is
\begin{equation}
 \psi(z) = c_{1}\,\sqrt{pz}\, J_{m}\bigl(2\sqrt{pz}\bigr) + c_{2}\,\sqrt{pz}\,Y_{m}\bigl(2\sqrt{pz}\bigr).
 \end{equation}

\noindent  Using standard properties of Bessel functions\footnote{In particular, it is useful to note $Y_{m}(x) = \frac{2}{\pi}J_{m}(x)\log \left(\frac{x}{2}\right) - \left(\frac{x}{2}\right)^{-|m|}P_{m}(x^{2})\,,
$  where $P_{m}(x^{2})$ is an analytic function of $x^{2}$ around $x=0\,$.} we can follow the solution as $z \to z e^{2\pi i}$ and from \eqref{eqn:monodef} we read off
\begin{eqnarray}\label{resonant monodromy}
M_{\gamma} =
 -e^{i\pi m}\left(
\begin{array}{cc}
1& 2i\frac{c_{2}}{c_{1}} \\
0 & 1
\end{array}
\right).
\end{eqnarray}

\noindent The difference with the non-resonant case is that the monodromy matrix has now only \textit{one} non-zero eigenvector, and it is a non-diagonalizable Jordan block. Note that the ratio $c_{2}/c_{1}$ would be fixed if we impose a boundary condition at $z=\infty$, say. Hence, the resonant monodromy depends on global properties of the solution, and not just local data in the vicinity of $z=0\,$. In particular, subleading terms in the expansion of $T(z)$ around $z=0$ are now important. Note however that the eigenvalues of the monodromy matrix, namely the local monodromy data, can be correctly obtained by the naive analytic continuation of the eigenvalues of the non-resonant monodromy matrix \eqref{non resonant monodromy} to integer $\nu\,$.

\bibliographystyle{uiuchept}
\bibliography{HigherSpin}

\end{document}